\documentclass[aps,twocolumn,nofootinbib,showpacs,prd,aps,10pt]{revtex4-1}
\usepackage[dvips]{graphicx}
\usepackage[english]{babel}
\usepackage[T1]{fontenc}
\usepackage{mathrsfs}
\usepackage[tbtags]{amsmath}
\usepackage{amssymb}
\usepackage{amsxtra}
\usepackage{amsopn}
\usepackage{latexsym}
\usepackage[mathcal]{eucal}
\usepackage{mathtools}

\newcommand{\BE}{\begin{equation}}
\newcommand{\EE}{\end{equation}}
\newcommand{\BA}{\begin{align}}
\newcommand{\EA}{\end{align}}

\newcommand{\nn}{\nonumber}

\newcommand{\tx}{\text}
\newcommand{\mc}{\mathcal}

\begin{document}

\title{One-loop RG improvement of the screened massive expansion in the Landau gauge}

\author{Giorgio Comitini}
\email{giorgio.comitini@dfa.unict.it}
\author{Fabio Siringo}
\email{fabio.siringo@ct.infn.it}

\affiliation{Dipartimento di Fisica e Astronomia 
dell'Universit\`a di Catania,\\ 
INFN Sezione di Catania,
Via S. Sofia 64, I-95123 Catania, Italy}

\date{\today}

\begin{abstract}
The RG improvement of the screened massive expansion is studied at one loop in two renormalization schemes, the momentum subtraction (MOM) scheme and the screened momentum subtraction (SMOM) scheme. The respective Taylor-scheme running couplings are shown not to develop a Landau pole, provided that the initial value of the coupling is sufficiently small. The improved ghost and gluon propagators are found to behave as expected, displaying dynamical mass generation for the gluons and the standard UV limit of ordinary perturbation theory. In the MOM scheme, when optimized by a matching with the fixed-coupling framework, the approach proves to be a powerful method for obtaining propagators which are in excellent agreement with the lattice data already at one loop. After optimization, the gluon mass parameter is left as the only free parameter of the theory and is shown to play the same role of the ordinary perturbative QCD scale $\Lambda_{\tx{QCD}}$.
\end{abstract}




\maketitle
\section{Introduction}

Being able to describe the non-perturbative regime of QCD is of paramount importance for understanding the low-energy phenomenology of hadrons, for predicting the observed hadron-mass spectrum and for addressing many unsolved problems like confinement, chiral symmetry breaking and dynamical mass generation \cite{cornwall,bernard,dono,philip,aguilar04,papa15b,olive}. 
Indeed, almost all of the observed mass in the universe seems to be generated by such mechanisms. Unfortunately, since perturbation theory (PT) breaks down in the infrared of QCD and of the pure-gauge Yang-Mills (YM) theory, to date a complete analytical treatment of the non-perturbative low-energy regime is still missing. In the last decades a considerable amount of knowledge
has been provided by numerical methods based on lattice calculations \cite{olive,cucch07,cucch08,cucch08b,cucch09,bogolubsky,olive09,dudal,binosi12,olive12,burgio15,duarte}
and numerical integration of integral equations in the continuum \cite{boucaud,boucaud2,aguilar8,aguilar10,aguilar14,rodriguez,rodriguez2,papa15,fischer2009,
huber14,huber15g,huber15b,pawlowski08,pawlowski10,pawlowski10b,pawlowski13,
varqcd,genself,highord,watson10,watson12,rojas,reinhardt04,reinhardt05,reinhardt14}. The breakdown of PT and the lack of an alternative analytical approach from first principles has also motivated the study of phenomenological models, mainly based on {\it ad hoc} modified Lagrangians~\cite{GZ,dudal08,dudal08b,dudal11,tissier10,tissier11,serreau}.

In the last years, a purely analytical approach to the exact gauge-fixed Lagrangian of QCD has been developed \cite{ptqcd,ptqcd2,scaling,analyt,xigauge,damp,varT,xighost,beta} based on a mere change of the expansion point of ordinary PT, showing that the breakdown of the theory may not be due to the perturbative method itself, but rather a consequence of a bad choice of its zero-order Lagrangian -- namely that of a massless free-particle theory --, which is good enough only in the UV because of asymptotic freedom. In the IR, because of mass generation, a {\it massive} free-particle theory could constitute the best expansion point, leading to a screened perturbative expansion which does not break down at any energy scale and is under control if the coupling is moderately small (as it turns out to be). Then, quite paradoxically, the {\it non-perturbative} regime of QCD and YM theory may be accessible by plain PT. Furthermore, in the IR and as far as the two-point functions are concerned, the higher-order terms of the perturbative series were shown to be minimized by an optimal choice of the renormalization scheme \cite{xigauge,xighost,beta}, yielding a very predictive analytical tool and one-loop results that are in excellent agreement with the available lattice data for YM theory. A remarkable feature of this {\it optimized} expansion is that the method is genuinely from first principles and does not require any external input apart from fixing the energy units.

The screened massive expansion shares with ordinary PT the problem of large logs that limit the validity of the optimized expansion to a low energy range, up to about 2 GeV \cite{beta}. In this paper we show how the problem can be solved by the Renormalization Group (RG), yielding an improved screened expansion whose validity can be virtually extended to any energy scale. Our findings corroborate the idea that QCD is a complete theory valid at all energies. In what follows, the RG-improved screened expansion is studied at one loop for the pure-gauge YM theory in two different renormalization schemes, and is shown to be under control down to arbitrarily small scales, even if higher-order terms become important in the IR, where the one-loop RG-improved results get worse than the optimized fixed-coupling expressions. Eventually, a matching between the two expansions provides a good agreement with the lattice data at all energies.

It is remarkable that, at one loop, the RG equation for the coupling can be integrated exactly in the different schemes, providing analytical expressions for the running coupling which merge with the universal one-loop result in the UV. In the IR, due to the non-perturbative scale set by the gluon mass, the coupling is scheme-dependent and finite if the flow starts from a moderate value in the UV, smaller than a threshold value. Above that threshold the running coupling develops an IR Landau pole.

This paper is organized as follows. In Sec. II the optimized screened expansion is reviewed for pure YM theory and its general renormalization and RG improvement are discussed. In Sec. III the RG-improved expansion is studied in the momentum-subtraction (MOM) scheme and in its screened version, which we term screened-MOM (SMOM). In Sec. IV the results of the previous sections are compared with the predictions of the optimized fixed-scale expansion and with the available lattice data. A matching between the two expansions provides a predictive theory which is in good agreement with the lattice data at all energy scales. Finally, in Sec.~V the main results are summarized and discussed.

\section{The screened massive expansion and its renormalization in the Landau gauge}

The screened {\it massive} expansion for the gauge-fixed and renormalized YM Lagrangian was first developed in Refs.~\cite{ptqcd,ptqcd2}, and extended to finite temperature in Refs.~\cite{damp,varT} and to the full QCD in Ref.~\cite{analyt}. The extension to a generic covariant gauge \cite{xigauge,xighost} has already demonstrated the predictive power of the method when the expansion is optimized by the constraints of the Becchi-Rouet-Stora-Tyutin (BRST) symmetry satisfied by the Faddeev-Popov Lagrangian. The renormalization of the screened expansion in the Landau gauge was discussed in Ref.~\cite{beta}, where different renormalization schemes were considered and analytical expressions were reported for the beta function.

The screened expansion is obtained by a shift of the expansion point of PT, performed {\it after} having renormalized the fields and the coupling, as discussed in Ref.~\cite{beta}. Following Refs.~\cite{ptqcd2,xigauge}, the shift is enforced by simply adding a transverse mass term to the quadratic part of the action and subtracting it again from the interaction, so that the total action is left unchanged. The action term which is added and subtracted is given by
\BE
\delta S= \frac{1}{2}\int A_{a\mu}(x)\>\delta_{ab}\> \delta\Gamma^{\mu\nu}(x,y)\>
A_{b\nu}(y) {\rm d}^4\, x{\rm d}^4y,
\label{dS1}
\EE
where the vertex function $\delta\Gamma$ is a shift of the inverse propagator,
\BE
\delta \Gamma^{\mu\nu}(x,y)=
\left[{\Delta_m^{-1}}^{\mu\nu}(x,y)- {\Delta_0^{-1}}^{\mu\nu}(x,y)\right],
\label{dG}
\EE
and $\Delta_{m}^{\mu\nu}$ is a massive free-particle propagator,
\begin{align}
{\Delta_m^{-1}}^{\mu\nu} (p)&=
(-p^2+m^2)\,t^{\mu\nu}(p)  
+\frac{-p^2}{\xi}\ell^{\mu\nu}(p),
\label{Deltam}
\end{align}
with the transverse and longitudinal projectors defined according to
\BE
t_{\mu\nu} (p)=g_{\mu\nu}  - \frac{p_\mu p_\nu}{p^2}\ ,\quad
\ell_{\mu\nu} (p)=\frac{p_\mu p_\nu}{p^2}.
\label{tl}
\EE
Adding the term $\delta S$ is equivalent to substituting the new massive propagator $\Delta_{m}^{\mu\nu}$ for the old massless one $\Delta_{0}^{\mu\nu}$ in the quadratic part of the action. The shift itself is motivated a posteriori by the former being much closer to the exact propagator in the IR than the latter, and a priori by a Gaussian Effective Potential (GEP) analysis of pure YM theory \cite{varT}.

In order to leave the total action unchanged, the opposite term $-\delta S$ is added in the interaction, providing a new two-point interaction vertex $\delta\Gamma$. Dropping all color indices in the diagonal matrices and inserting Eq.~(\ref{Deltam}) in Eq.~(\ref{dG}), the vertex is just the transverse mass shift of the quadratic part,
\BE
\delta \Gamma^{\mu\nu} (p)=m^2 t^{\mu\nu}(p).
\label{dG2}
\EE
The new vertex does not contain any renormalization constant and is part of the interaction even if it does not explicitly depend on the coupling. Thus the expansion itself must be regarded as a $\delta$-expansion, rather than a loop expansion, since different powers of the coupling coexist at each order in powers of the total interaction.

The self-energies and the propagators are evaluated, order by order, by PT, with a modified set of Feynman rules by which the gluon lines are associated to massive free-particle propagators $\Delta_{m}^{\mu\nu}$ and the new two-point vertex $\delta \Gamma^{\mu\nu}$ is included in the graphs. Since the total gauge-fixed Faddeev-Popov Lagrangian is not modified and because of gauge invariance, the exact gluon longitudinal polarization is known to vanish. The exact gluon polarization can thus be written as
\BE
\Pi^{\mu\nu}(p)=\Pi(p^{2})\, t^{\mu\nu}(p).
\label{pol}
\EE
It follows that in the Landau gauge, $\xi=0$, the exact gluon propagator is transverse,
\BE
\Delta_{\mu\nu}(p)=\Delta (p^{2})\,t_{\mu\nu}(p),
\EE
and defined by the single scalar function $\Delta(p^{2})$. In the Euclidean formalism and Landau gauge, the dressed gluon and ghost propagators of the screened expansion can be expressed as
\begin{align}
\nn\Delta^{-1}(p^{2})&=p^{2}+m^{2}-\Pi(p^{2}),\\
\mc{G}^{-1}(p^{2})&=-p^{2}-\Sigma(p^{2}),
\label{dressprop}
\end{align}
where the proper gluon polarization $\Pi(p^{2})$ and ghost self-energy $\Sigma(p^{2})$ are the sum of all one-particle-irreducible (1PI) graphs in the screened expansion, including the mass and renormalization counterterms.

It is important to keep in mind that, since the total Lagrangian is not modified, the exact renormalization constants satisfy the Slavnov-Taylor identities. Nonetheless, the added mass term breaks the BRST symmetry of the quadratic part and of the interaction when these are taken apart. Therefore, some of the constraints arising from BRST symmetry are not satisfied exactly at any finite order of the screened expansion. While the soft breaking has no effect on the UV behavior and on the diverging parts of the renormalization constants, some spurious diverging mass terms do appear in the expansion at some stage. However, as discussed in Refs.~\cite{ptqcd,ptqcd2,analyt,xigauge}, the insertions of the new vertex $\delta \Gamma$, Eq.~(\ref{dG2}), cancel the spurious divergences exactly, without the need of any mass renormalization counterterm, as a consequence of the unbroken BRST symmetry of the whole action. This aspect makes the screened expansion very different from effective models where a bare mass term is added to the Lagrangian from the beginning. In the screened massive expansion, the gluon mass parameter is an arbitrary and finite quantity which is added and subtracted again in the renormalized action and, as such, it can be taken to be an RG invariant.

As shown for instance in Ref.~\cite{ptqcd2}, the exact self-energies of the screened expansion can be written as
\begin{align}
\nn\Pi(p^{2})&=m^{2}-p^{2}\delta Z_{A}+\Pi_{\tx{loop}}(p^{2}),\\
\Sigma(p^{2})&=p^{2}\delta Z_{c}+\Sigma_{\tx{loop}}(p^{2}),
\label{selfs0}
\end{align}
where the tree-level contribution $m^{2}$ comes from the new two-point vertex $\delta \Gamma$ in Eq.~\eqref{dG2}, while the tree-level terms $-p^{2}\delta Z_A$, $p^{2}\delta Z_c$ arise from the respective field-strength renormalization counterterms. Observe that the vertex mass term in Eq.~\eqref{selfs0} exactly cancels the zero-order gluon propagator's mass in Eq.~\eqref{dressprop}: in the screened expansion, the gluon's mass is not a mere artifact of the choice of a massive tree-level propagator, but rather it is dynamically generated by the loops' contribution to the self-energy (more precisely, it comes from the gluon loops \cite{ptqcd,ptqcd2,xigauge}). Indeed, the screened expansion of QED would not predict the existence of a mass for the photons, which are not self-interacting.

The proper functions $\Pi_{\tx{loop}}(p^{2})$, $\Sigma_{\tx{loop}}(p^{2})$ are given by the sum of all 1PI graphs containing loops. The diverging parts of $\delta Z_{A}$, $\delta Z_{c}$ cancel the UV divergences of $\Pi_{\tx{loop}}$ and $\Sigma_{\tx{loop}}$, respectively. Since these divergences do not depend on mass scales, they are exactly the same as in the standard PT, so that in the $\overline{MS}$ scheme $Z_A$ and $Z_c$ have their standard expressions, as manifest in the explicit one-loop calculation~\cite{ptqcd,ptqcd2,beta}. The finite parts of $\delta Z_A$, $\delta Z_c$, on the other hand, are arbitrary and depend on the renormalization scheme. Indeed, the self-energies themselves each contain an arbitrary term of the form ${\cal C}p^2$, where ${\cal C}$ is a constant whose value depends on the regularization method.

\begin{figure}[b]
\vskip 1cm
\centering
\includegraphics[width=0.23\textwidth,angle=-90]{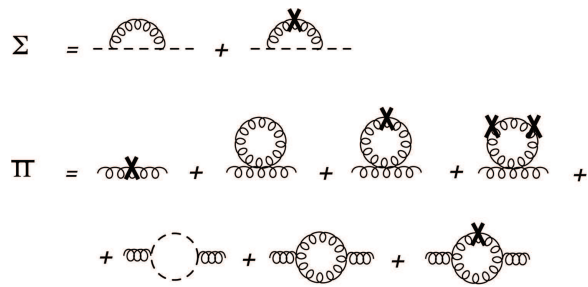}
\caption{Diagrams that contribute to the ghost self-energy and gluon polarization to third order in the $\delta$-expansion and one loop. The crosses denote the insertions of the vertex $\delta\Gamma$.}
\label{figdiagrams}
\end{figure}

To one loop, the explicit expressions for the loop self-energies, as computed from the diagrams in Fig.~\ref{figdiagrams}, can be written as
\begin{align}
\nn\Pi_{\tx{loop}}(p^{2})&=\alpha p^{2}\left\{\frac{13}{18}\left(\frac{2}{\epsilon}+\ln\frac{\overline{\mu}^{2}}{m^{2}}\right)-F(s)-\mc{C}\right\},\\
\Sigma_{\tx{loop}}(p^{2})&=-\alpha p^{2}\left\{\frac{1}{4}\left(\frac{2}{\epsilon}+\ln\frac{\overline{\mu}^{2}}{m^{2}}\right)-G(s)-\mc{C}'\right\},
\label{selfs}
\end{align}
where
\BE
\alpha=\frac{3N\alpha_{s}}{4\pi}=\frac{3Ng^{2}}{16\pi^{2}},
\EE
$\mc{C}$ and $\mc{C}'$ are constants and $F(s)$, $G(s)$ are dimensionless functions of the ratio $s=p^2/m^2$, whose explicit expressions were derived in Refs.~\cite{ptqcd,ptqcd2} and are reported in the Appendix. For further details on the screened expansion we refer to~\cite{xigauge,xighost,beta}, where explicit analytical expressions for the propagators are reported to third order in the $\delta$-expansion and to one loop, also in an arbitrary covariant gauge.\\

While the exact observables must be RG-invariant and cannot depend on the renormalization scale, the approximate one-loop expressions do depend on the scale and on the scheme. Moreover, some exact consequences of BRST symmetry, like the Nielsen identities \cite{nielsen,kobes90,breck}, might not be satisfied at any finite order of the screened expansion. An optimal choice of the finite parts of the renormalization constants provides propagators which are closer to the exact, RG-invariant result, and can be determined by the principle of minimal sensitivity \cite{sensitivity}. The resulting optimized PT is known as renormalization-scheme optimized PT \cite{stevensonRS} and turns out to be quite effective.

For an observable particle, the finite parts are usually fixed on mass shell. For instance, the Nielsen identities are satisfied at any finite order of PT for electrons and quarks when the self energy is renormalized on shell \cite{breck}. For the gluons, without an observable mass at hand,  the argument can be reversed. The scheme can be defined by imposing that the Nielsen identities are satisfied, i.e. by requiring that the poles and residues of the propagator be gauge-parameter independent. While this condition is not generally satisfied at one loop, in Refs.~\cite{xigauge,beta} we showed that there exists an optimal choice of the renormalization constants which makes the pole structure gauge invariant. For this special choice the higher-order terms turn out to be minimal and negligible in the IR, so that the optimized one-loop analytical expressions provide an excellent agreement with the available low-energy lattice data when the energy scale is fixed by setting $m=0.656$ GeV. The resulting optimized expansion is very predictive and gives valuable quantitative information on the analytical properties in Minkowski space even for different covariant gauges, which are not accessible by lattice calculations.

Unfortunately, being based on an optimal choice of the renormalization scale, the optimized expansion is not reliable for $p/m\gtrsim 3$ (corresponding to $p\gtrsim 2$ GeV for $m=0.656$ GeV) because of the large logs. For instance, in Eq.~(\ref{selfs}), the ghost self energy contains a leading term $G(s)\approx \ln(s)/4$ which spoils the multiplicative renormalizability of the propagator for a finite change of scale, unless the shift $\mu^\prime-\mu\ll m$. This problem is usually solved by integrating the RG flow, yielding an improved version of the perturbative expansion.

The evaluation of the RG-improved gluon and ghost propagators requires the knowledge of the respective anomalous dimensions and of the beta function. In a momentum-subtraction-like renormalization scheme defined by the values of the propagators and coupling at the scale $\mu$, the calculation of the anomalous dimensions and beta function from the explicit expressions of the self energies in Eqs.~(\ref{selfs}) is straightforward. At $p^{2}=\mu^{2}$, using Eqs.~\eqref{dressprop}-\eqref{selfs0}, we can write
\begin{align}
\nn\mu^{-2}\Delta^{-1}(\mu^{2})&=1+\delta Z_{A}-\mu^{-2}\,\Pi_{\tx{loop}}(\mu^{2}),\\
-\mu^{-2}\mc{G}^{-1}(\mu^{2})&=1+\delta Z_{c}+\mu^{-2}\,\Sigma_{\tx{loop}}(\mu^{2}),
\end{align}
so that
\begin{align}\label{fieldstrength}
\nn Z_{A}&=\mu^{-2}\left[\Delta^{-1}(\mu^{2})+\Pi_{\tx{loop}}(\mu^{2})\right],\\
Z_{c}&=-\mu^{-2}\left[\mc{G}^{-1}(\mu^{2})+\Sigma_{\tx{loop}}(\mu^{2})\right].
\end{align}
The gluon and ghost anomalous dimensions $\gamma_{A}$ and $\gamma_{c}$ are then defined as
\begin{align}\label{anom}
\gamma_{A}=\frac{1}{2}\frac{d\ln Z_{A}}{d\ln\mu}\ ,\qquad\quad \gamma_{c}=\frac{1}{2}\frac{d\ln Z_{c}}{d\ln\mu}.
\end{align}
As for the renormalized strong coupling constant $g$, this can be defined as
\BE
g=g_{B}\,\frac{Z_{c}Z_{A}^{1/2}}{Z_{1}^{c}},
\EE
where $g_{B}$ is the bare coupling and $Z_{1}^{c}$ is the renormalization factor of the ghost-gluon vertex. In the Landau gauge, $\xi=0$, the divergent part of the ghost-gluon vertex is known to vanish, so that $Z_{1}^{c}$ is finite. The simplest renormalization condition for the vertex is therefore given by $Z_{1}^{c}=1$. The latter defines the Taylor scheme \cite{taylor,vonsmek,stern,boucaud3}, in which
\BE\label{couptaylor}
g=g_{B}\,Z_{c}Z_{A}^{1/2}.
\EE
From the above equation we can immediately derive the beta function:
\BE\label{beta}
\beta=\mu\frac{dg}{d\mu}=g(2\gamma_{c}+\gamma_{A}).
\EE
Thus in the Taylor scheme the knowledge of $\gamma_{A}$ and $\gamma_{c}$ is sufficient for computing $\beta$.

The RG-improved propagators renormalized at the scale $\mu_{0}$ are defined in terms of the anomalous dimensions according to
\begin{align}\label{rgprop}
\nn\Delta(p^{2};\mu_{0})&=\widehat{\Delta}(p^{2})\exp\left(\int_{\mu_{0}^{2}}^{p^{2}}\frac{d\mu^{\prime\,2}}{\mu^{\prime\,2}}\,\gamma_{A}(\mu^{\prime\,2})\right),\\
\mc{G}(p^{2};\mu_{0})&=\widehat{\mc{G}}(p^{2})\exp\left(\int_{\mu_{0}^{2}}^{p^{2}}\frac{d\mu^{\prime\,2}}{\mu^{\prime\,2}}\,\gamma_{c}(\mu^{\prime\,2})\right).
\end{align}
Here $\widehat{\Delta}(p^{2})$ and $\widehat{\mc{G}}(p^{2})$ are scheme-dependent functions that are determined by the renormalization conditions: since for any value of the initial renormalization scale
\begin{align}\label{rgpref}
\nn\widehat{\Delta}(\mu^{2}_{0})&=\Delta(\mu_{0}^{2};\mu_{0}),\\
\widehat{\mc{G}}(\mu_{0}^{2})&=\mc{G}(\mu_{0}^{2};\mu_{0}),
\end{align}
the functions $\widehat{\Delta}$, $\widehat{\mc{G}}$ evaluated at $p^{2}$ are simply equal to the values of the respective propagators, renormalized at $\mu^{2}=p^{2}$ and evaluated at the same scale.\\

In the next section we will investigate the behavior of the one-loop RG-improved propagators and running coupling in two renormalization schemes: the ordinary momentum subtraction (MOM) scheme and the screened momentum subtraction (SMOM) scheme. In the UV, any RG-improvement of the screened expansion must lead to the standard PT RG-improved results, since for $p\gg m$ the mass effects become irrelevant. It follows that the improved screened expansion predicts the correct asymptotic UV behavior for the propagators and coupling already at one loop. On the other hand, in the IR, where the one-loop optimized fixed-scale expansion of Refs.~\cite{xigauge,beta} has already proven successful, the RG-improved results may actually turn out to be quantitatively inaccurate (regardless of the value of $m$) when truncated to leading order: while the higher-order terms are minimal at the optimal scale, as the scale runs down with the momentum the higher-loop corrections to the anomalous dimensions can become quite large, since in the IR the running coupling becomes of order unity. Nevertheless, perhaps remarkably, it turns out that already at one loop the improvement of the screened expansion provides a qualitatively accurate picture of the IR behavior of the propagators, with a running coupling that does not exhibit a Landau pole. Quantitatively, we expect the accuracy of the approximation to improve by including the higher-order corrections to the anomalous dimensions and beta function.

The screened massive expansion introduces the gluon mass parameter $m$ as a spurious free parameter, whose value cannot be determined from first principles since Yang-Mills theory is scale-invariant at the classical level. Of course, the arbitrariness of $m$ results in a loss of predictivity of the method, allowing for infinitely many solutions for the YM $n$-point functions; namely, one for every pair $(m^{2},\alpha_{s}(\mu_{0}^{2}))$. In Sec.~III we do not address this issue; instead, we study the behavior of the gluon and ghost two-point functions by expressing every dimensionful quantity in units of $m$ and letting $\alpha_{s}(\mu_{0}^{2})$ vary. When needed for comparison, we will take $m=0.656$ GeV, as determined e.g. in Ref.~\cite{xigauge} by fitting the fixed-scale gluon propagator to the lattice data of Ref.~\cite{duarte}. Then, in Sec.~IV, we will present a method for optimizing the initial value of the coupling $\alpha_{s}(\mu_{0}^{2})$; the dimensionful value of the renormalization scale $\mu_{0}$ itself will depend on the mass scale set by $m$. With $\alpha_{s}(\mu_{0}^{2})$ fixed by optimization, the redundancy in the choice of free parameters is removed -- thus restoring the predictivity of the screened expansion -- and $m$ is left as the only free parameter to determine the physics of the theory, playing the same role of $\Lambda_{\tx{QCD}}$ in standard perturbation theory as the fundamental energy scale of YM theory.

\section{Running coupling and RG-improved propagators}
\subsection{MOM scheme}

The momentum subtraction (MOM) scheme is defined by the renormalization conditions
\begin{align}\label{condmom}
\nn\Delta^{-1}(\mu^{2})&=\mu^{2},\\
\mc{G}^{-1}(\mu^{2})&=-\mu^{2}.
\end{align}
When plugged into Eq.~\eqref{fieldstrength}, these lead to the following one-loop field strength renormalization counterterms (modulo irrelevant constants):
\begin{align}\label{countmom}
\nn \delta Z_{A}^{(\tx{MOM})}&=\alpha\left\{\frac{13}{18}\left(\frac{2}{\epsilon}+\ln\frac{\overline{\mu}^{2}}{m^{2}}\right)-F\left(\frac{\mu^{2}}{m^{2}}\right)\right\},\\
\delta Z_{c}^{(\tx{MOM})}&=\alpha\left\{\frac{1}{4}\left(\frac{2}{\epsilon}+\ln\frac{\overline{\mu}^{2}}{m^{2}}\right)-G\left(\frac{\mu^{2}}{m^{2}}\right)\right\}.
\end{align}
In the limit of large renormalization scales ($\mu^{2}\gg m^{2}$, $x\to \infty$),
\begin{align}\label{functhighlim}
\nn F(x)&\to \frac{13}{18}\,\ln x,\\
G(x)&\to \frac{1}{4}\,\ln x
\end{align}
(cf. the Appendix), and we recover the leading-order counterterms of ordinary PT. From Eq.~\eqref{countmom}, the one-loop gluon and ghost field anomalous dimensions in the MOM scheme follow as
\begin{align}\label{anommom}
\nn\gamma_{A}^{(\tx{MOM})}(\mu^{2})&=-\alpha(\mu^{2})\,\frac{\mu^{2}}{m^{2}}\ F'(\mu^{2}/m^{2}),\\
\gamma_{c}^{(\tx{MOM})}(\mu^{2})&=-\alpha(\mu^{2})\,\frac{\mu^{2}}{m^{2}}\ G'(\mu^{2}/m^{2}).
\end{align}
Due to the presence of the mass scale set by the gluon mass parameter $m$, the anomalous dimensions $\gamma_{A}^{(\tx{MOM})}$ and $\gamma_{c}^{(\tx{MOM})}$ depend explicitly on the renormalization scale, rather than only implicitly through the running coupling $\alpha(\mu^{2})$. This dependence is lost at high renormalization scales, where $F'(x)$ and $G'(x)$ are proportional to $x^{-1}$ (see Eq.~\eqref{functhighlim}) and the anomalous dimensions of ordinary PT are recovered.

\begin{figure}[b]
\vskip 1cm
\centering
\includegraphics[width=0.30\textwidth,angle=-90]{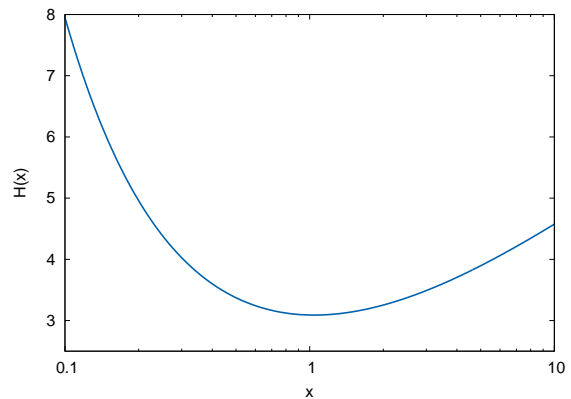}
\caption{Function $H(x)$. The minimum $H(x_{0})\approx 3.090$ is found at $x_{0}\approx1.044$.}
\label{fighfunct}
\end{figure}

To the coupling $\alpha$ we may associate a beta function $\beta_{\alpha}$, defined as
\BE\label{beta2}
\beta_{\alpha}=\frac{d\alpha}{d\ln\mu^{2}}=\alpha\,\frac{\beta}{g}.
\EE
Using Eq.~\eqref{beta}, $\beta_{\alpha}$ can be computed in the MOM scheme from the anomalous dimensions $\gamma_{A}^{(\tx{MOM})}$ and $\gamma_{c}^{(\tx{MOM})}$, yielding
\BE\label{betamom}
\beta_{\alpha}^{(\tx{MOM})}(\mu^{2})=-\alpha^{2}\,\frac{\mu^{2}}{m^{2}}\ H'(\mu^{2}/m^{2})
\EE
to one loop. Here the function $H(x)$, shown in Fig.~\ref{fighfunct}, is defined as
\BE
H(x)=2G(x)+F(x),
\EE
and has limiting behavior (see Eq.~\eqref{functhighlim})
\BE\label{hlim}
H(x)\to\frac{11}{9}\,\ln x\qquad\qquad(x\to\infty).
\EE
From Eq.~\eqref{betamom} we see that, along with the anomalous dimensions, the MOM beta function of the screened expansion also has an explicit dependence on the renormalization scale $\mu$. As we will show in a moment, this is a most important feature of the modified perturbation theory, bringing in mass effects which are able to prevent the developing of a Landau pole in the running coupling.\\

To one loop, the differential equation for the running coupling $\alpha^{(\tx{MOM})}(\mu^{2})$,
\BE
\frac{d\alpha^{(\tx{MOM})}}{d\ln s}=-(\alpha^{(\tx{MOM})})^{2}\ s\,H'(s),
\EE
($s=\mu^{2}/m^{2}$) can be solved exactly. In terms of $\alpha_{s}$, its solution is given by
\BE\label{coupmom}
\alpha^{(\tx{MOM})}_{s}(\mu^{2})=\frac{\alpha^{(\tx{MOM})}_{s}(\mu_{0}^{2})}{1+\frac{3N}{4\pi}\alpha_{s}^{(\tx{MOM})}(\mu_{0}^{2})\left[H(s)-H(s_{0})\right]},
\EE
where $\mu_{0}$ is the initial renormalization scale, $s_{0}=\mu_{0}^{2}/m^{2}$ and $\alpha_{s}^{(\tx{MOM})}(\mu^{2}_{0})$ is the value of the MOM coupling renormalized at $\mu_{0}$ (initial condition of the RG flow). This result was already derived directly from Eq.~\eqref{couptaylor} in Refs.~\cite{ptqcd,ptqcd2}.

\begin{figure}[t]
\vskip 1cm
\centering
\includegraphics[width=0.30\textwidth,angle=-90]{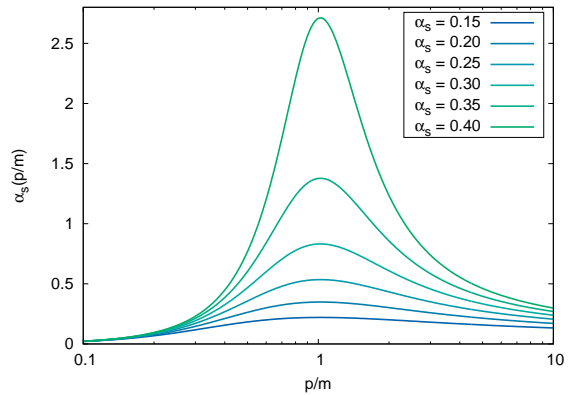}
\caption{$N=3$ one-loop running coupling of the screened expansion in the MOM scheme for different initial values of the coupling at the scale $\mu_{0}/m=6.098$. With $m=0.656$ GeV as in our previous works, this corresponds to $\mu_{0}=4$ GeV. The running coupling develops a Landau pole for $\alpha_{s}^{(\tx{MOM})}(\mu_{0}^{2})\geq 0.469$.}
\label{figcoupmom}
\end{figure}

In the limit of high initial and final renormalization scales ($s,s_{0}\gg 1$), using Eq.~\eqref{hlim}, it is easy to see that $\alpha_{s}^{(\tx{MOM})}(\mu^{2})$ reduces to the standard one-loop running coupling,
\BE\label{coupord}
\alpha_{s}^{(\tx{MOM})}(\mu^{2})\to\frac{\alpha_{s}(\mu_{0}^{2})}{1+\frac{11N}{3}\frac{\alpha_{s}(\mu_{0}^{2})}{4\pi}\ln(\mu^{2}/\mu^{2}_{0})}.
\EE
At intermediate and low momenta, on the other hand, the behavior of $\alpha_{s}^{(\tx{MOM})}(\mu^{2})$ radically differs from that of its counterpart in ordinary PT (see Fig.~\ref{figcoupmom}). Due to the explicit dependence of $\beta_{\alpha}^{(\tx{MOM})}$ on the renormalization scale, the latter is allowed to vanish already at one loop for a non-zero value of the coupling constant. The vanishing occurs at the fixed renormalization scale $\mu_{\star}$ that solves the equation
\BE
H'(\mu^{2}_{\star}/m^{2})=0.
\EE
Numerically, one finds that
\BE
\mu_{\star}\approx 1.022\ m
\EE
or $\mu_{\star}\approx 0.67$ GeV for $m=0.656$ GeV. Of course, since the beta function vanishes as a function of $\mu$, rather than for some specific value of the coupling, the existence of a zero for $\beta_{\alpha}^{(\tx{MOM})}$ does not result in a fixed point of the RG flow. Instead, it provides a mechanism by which, at scales of the order of the gluon mass parameter, the running of the coupling is allowed to slow down, thus making it possible to prevent the developing of a Landau pole in $\alpha_{s}^{(\tx{MOM})}(\mu^{2})$. Indeed, since $\mu_{\star}^{2}/m^{2}$ is actually a minimum for $H(s)$,
\BE
H(s)\geq H(\mu_{\star}^{2}/m^{2})\approx 3.090,
\EE
Eq.~\eqref{coupmom} implies that the one-loop MOM running coupling remains finite at all renormalization scales, provided that its value renormalized at the scale $\mu_{0}$ is smaller than the scale-dependent threshold value $\alpha_{\tx{pole}}^{(\tx{MOM})}(\mu_{0}^{2})$ defined by
\BE\label{threshmom}
\alpha_{\tx{pole}}^{(\tx{MOM})}(\mu_{0}^{2})=\frac{1}{H(\mu_{0}^{2}/m^{2})-H(\mu_{\star}^{2}/m^{2})}.
\EE
At $\mu_{0}=6.098\, m$ (corresponding to $\mu_{0}=4$ GeV in physical units), Eq.~\eqref{threshmom} yields
\BE
\alpha_{\tx{pole}}^{(\tx{MOM})}(6.098\, m)\approx0.336,
\EE
or, in terms of $\alpha_{s}=4\pi\alpha/3N$,
\BE
\alpha_{s,\tx{pole}}^{(\tx{MOM})}(6.098\, m)\approx0.469
\EE
for $N=3$. If $\alpha^{(\tx{MOM})}(\mu_{0}^{2})\geq \alpha_{\tx{pole}}^{(\tx{MOM})}(\mu_{0}^{2})$, the denominator of Eq.~\eqref{coupmom} eventually vanishes and the running still encounters a Landau pole: for $\alpha^{(\tx{MOM})}(\mu_{0}^{2})=\alpha_{\tx{pole}}^{(\tx{MOM})}(\mu_{0}^{2})$ the pole is found exactly at $\mu=\mu_{\star}$, whereas for larger values of the coupling it is found at scales between $\mu_{\star}$ and $\mu_{0}$.

If the initial value of the coupling is smaller than $\alpha_{\tx{pole}}^{(\tx{MOM})}$, as the momentum decreases the one-loop running coupling remains finite and attains a maximum at $\mu=\mu_{\star}$, where the beta function switches from being negative to being positive and $\alpha_{s}^{(\tx{MOM})}(\mu^{2})$ starts to decrease. The value of the coupling at the maximum is an increasing and unbounded function of $\alpha_{s}^{(\tx{MOM})}(\mu^{2}_{0})$. At vanishing renormalization scales ($\mu^{2}\ll m^{2}$), due to the limiting behavior
\BE
H(x)\to\frac{5}{8x}\qquad\qquad(x\to 0)
\EE
(cf. the Appendix), the running coupling decreases linearly with $\mu^{2}$,
\BE
\alpha_{s}^{(\tx{MOM})}(\mu^{2})\to \frac{32\pi}{15N}\ \frac{\mu^{2}}{m^{2}},
\EE
and tends to zero with a derivative that does not depend on the initial conditions of the RG flow. As we will see, even if the coupling vanishes at $\mu=0$, the low-energy dynamics of the gluons remains highly non-trivial.\\

\begin{figure}[b]
\vskip 1cm
\centering
\includegraphics[width=0.30\textwidth,angle=-90]{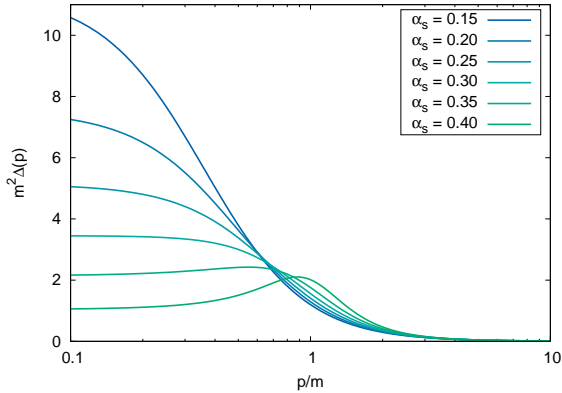}
\caption{$N=3$ one-loop RG-improved gluon propagator in the MOM scheme, renormalized at the scale $\mu_{0}/m=6.098$ (corresponding to $\mu_{0}=4$ GeV for $m=0.656$ GeV), computed for different initial values of the coupling at the same scale.}
\label{figglumom}
\end{figure}

\begin{figure}[t]
\vskip 1cm
\centering
\includegraphics[width=0.30\textwidth,angle=-90]{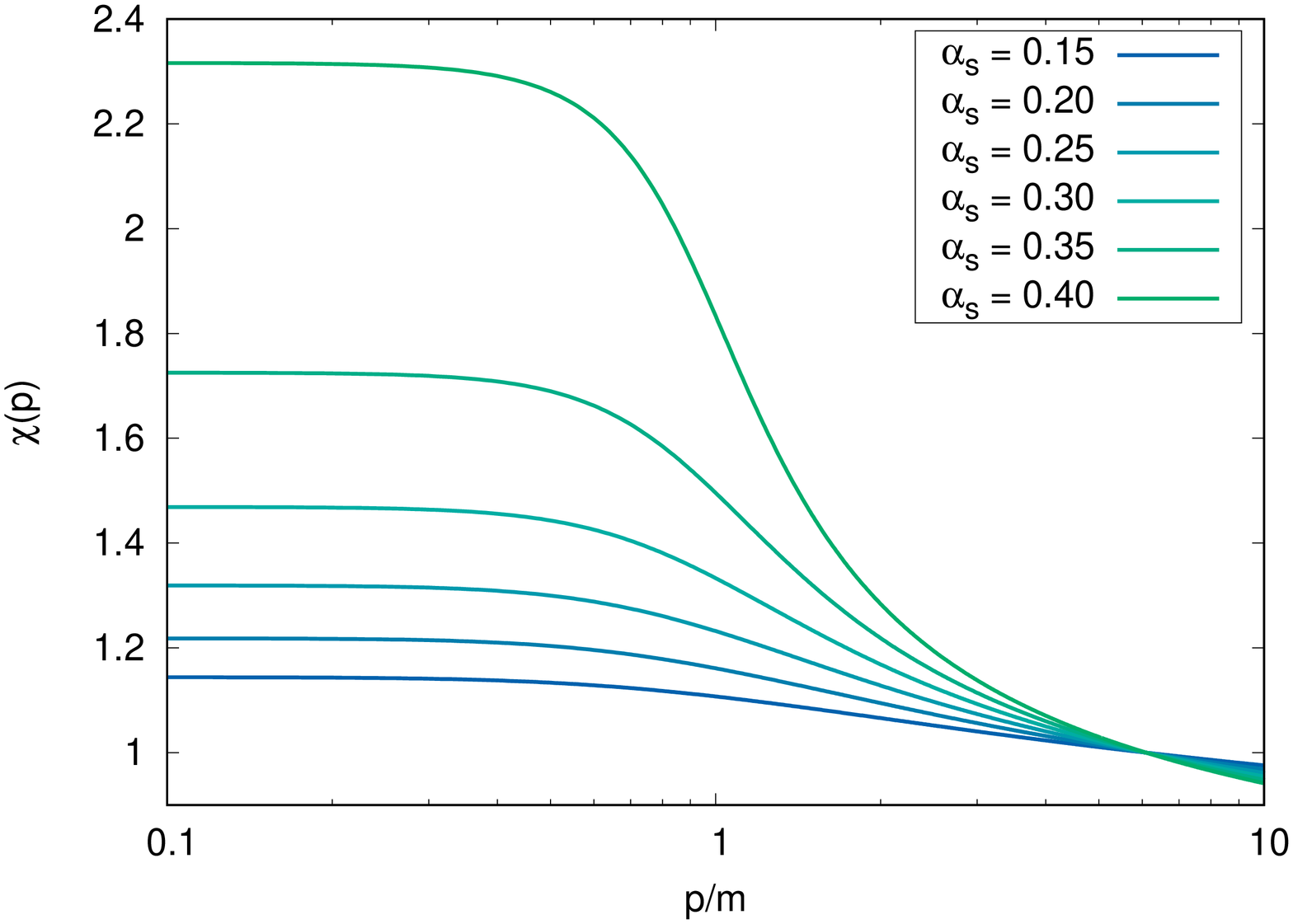}
\caption{$N=3$ one-loop RG-improved ghost dressing function $\chi(p)=-p^{2}\mc{G}(p)$ in the MOM scheme, renormalized at the scale $\mu_{0}/m=6.098$ (corresponding to $\mu_{0}=4$ GeV for $m=0.656$ GeV), computed for different initial values of the coupling at the same scale.}
\label{figghomom}
\end{figure}

Once the running coupling is known, the RG-improved gluon and ghost propagators can be computed using Eq.~\eqref{rgprop} by an appropriate choice of the functions $\widehat{\Delta}(p^{2})$ and $\widehat{\mc{G}}(p^{2})$. In the MOM scheme, in order to fulfill the renormalization conditions given by Eq.~\eqref{condmom}, one must set
\begin{align}
\nn\widehat{\Delta}^{(\tx{MOM})}(p^{2})&=\frac{1}{p^{2}},\\
\widehat{\mc{G}}^{(\tx{MOM})}(p^{2})&=-\frac{1}{p^{2}}
\end{align}
(see Eq.~\ref{rgpref}). The one-loop RG-improved propagators renormalized at the scale $\mu_{0}$ then read
\begin{widetext}
\begin{align}\label{rgpropmom}
\nn\Delta^{(\tx{MOM})}(p^{2};\mu_{0}^{2})&=\frac{1}{p^{2}}\exp\left(-\int_{\mu_{0}^{2}/m^{2}}^{p^{2}/m^{2}}ds\ \alpha^{(\tx{MOM})}(s)\,F'(s)\right),\\
\mc{G}^{(\tx{MOM})}(p^{2};\mu_{0}^{2})&=-\frac{1}{p^{2}}\exp\left(-\int_{\mu_{0}^{2}/m^{2}}^{p^{2}/m^{2}}ds\ \alpha^{(\tx{MOM})}(s)\,G'(s)\right),
\end{align}
\end{widetext}
where the running coupling is expressed as a function of the adimensional variable $s=\mu^{2}/m^{2}$. The one-loop improved gluon propagator and ghost dressing function renormalized at the scale $\mu_{0}=6.098\,m$ (corresponding to $\mu_{0}=4$ GeV in physical units) are shown respectively in Fig.~\ref{figglumom} and Fig.~\ref{figghomom} for different initial values of the coupling constant below the threshold value $\alpha_{s,\tx{pole}}^{(\tx{MOM})}\approx 0.47$.

Since in the high momentum limit the MOM anomalous dimensions and running coupling reduce to their standard one-loop perturbative expression, asymptotically\footnote{Provided that the initial renormalization scale $\mu_{0}$ is much larger than $m$.} the one-loop RG-improved propagators behave as known fractional powers of the running coupling divided by the momentum squared,
\begin{align}\label{stdrgprop}
\nn \Delta^{(\tx{MOM})}(p^{2})&\to\frac{1}{p^{2}}\left[\frac{\alpha_{s}(p^{2})}{\alpha_{s}(\mu_{0}^{2})}\right]^{13/22},\\
\mc{G}^{(\tx{MOM})}(p^{2})&\to-\frac{1}{p^{2}}\left[\frac{\alpha_{s}(p^{2})}{\alpha_{s}(\mu_{0}^{2})}\right]^{9/44}.
\end{align}
At intermediate and low momenta, if the running coupling does not develop a Landau pole, the one-loop improved gluon propagator attains a maximum at the momentum $p$ that solves the equation
\BE\label{glumaxmom}
1+\alpha^{(\tx{MOM})}(p^{2})\,\frac{p^{2}}{m^{2}}\,F'(p^{2}/m^{2})=0.
\EE
That Eq.~\eqref{glumaxmom} always admits a solution follows from the asymptotic behavior
\begin{align}
\nn1+\alpha^{(\tx{MOM})}(s)\,sF'(s)&\to\frac{2}{45}\,s\ln s\leq0\qquad(s\to 0),\\
1+\alpha^{(\tx{MOM})}(s)\, sF'(s)&\to1>0\qquad (s\to \infty)
\end{align}
(cf. the Appendix). The position of the maximum depends on the initial conditions of the running, and shifts from higher to lower momenta as $\alpha_{s}^{(\tx{MOM})}(\mu_{0}^{2})$ is decreased, eventually coming arbitrarily close to $p=0$. At vanishingly small momenta, due to the low energy limits
\begin{align}
\nn \alpha^{(\tx{MOM})}(s)F'(s)&\to-\frac{1}{s},\\
\alpha^{(\tx{MOM})}(s)G'(s)&\to-\frac{4}{15}s\ln s
\end{align}
(cf. the Appendix), the one-loop improved propagators behave as
\begin{align}\label{propmomlow}
\nn\Delta^{(\tx{MOM})}(p^{2})&\to\frac{s\,e^{k}}{p^{2}}=\frac{e^{k}}{m^{2}},\\
\mc{G}^{(\tx{MOM})}(p^{2})&\to-\frac{e^{k'}}{p^{2}},
\end{align}
where $k$ and $k'$ are constants that generally depend on the initial conditions of the running. Since $\Delta^{(\tx{MOM})}(p^{2})$ remains finite as $p^{2}\to 0$, in the MOM-scheme RG-improved picture the gluons are still predicted to dynamically acquire a mass. The ghosts, on the other hand, remain massless ($\mc{G}^{(\tx{MOM})}(p^{2})\to \infty$ as $p^{2}\to 0$).\\

The most notable feature of the one-loop RG-improved screened expansion in the MOM scheme is the absence of a Landau pole in its running coupling for sufficiently small initial values of $\alpha_{s}^{(\tx{MOM})}(\mu_{0}^{2})$, a necessary condition for the consistency of a perturbation theory which aims to be valid at all energy scales. As we saw, instead of growing to infinity at a finite momentum, the one-loop MOM coupling interpolates between the standard high-energy logarithmic behavior and a decreasing low-energy behavior ($\alpha_{s}^{(\tx{MOM})}(p^{2})\sim p^{2}$ as $p^{2}\to 0$) by attaining a maximum at the fixed scale $\mu_{\star}\approx1.022\ m$. Depending on the initial conditions of the RG flow, the value of the coupling at the maximum can become quite large for the perturbative standards. As a consequence, the higher orders of the perturbative expansion might become significant at scales comparable to that of the gluon mass parameter.

Since our one-loop, low-energy results evolve from a region of generally large couplings, we should expect these to give, at best, a good qualitative approximation of the exact, non-perturbative behavior of Yang-Mills theory for any given value of the pair $(m^{2},\alpha_{s}(\mu_{0}^{2}))$. In the absence of estimates for the higher-order corrections to the propagators, the extent to which the approximation is good can be established only a posteriori, by a comparison with non-perturbative results such as those obtained on the lattice. This aspect will be investigated in Sec.~IV, where we will also propose a method for fixing the value of the spurious free parameter (either the gluon mass parameter $m$ or the value of the coupling at some fixed renormalization scale) of the RG-improved screened expansion.

\subsection{SMOM scheme}

The screened momentum subtraction (SMOM) scheme~\cite{beta} is defined by the renormalization conditions
\begin{align}\label{condsmom}
\nn\Delta^{-1}(\mu^{2})&=\mu^{2}+m^{2},\\
\mc{G}^{-1}(\mu^{2})&=-\mu^{2}.
\end{align}
To one loop, these require the field strength counterterms to be chosen (modulo irrelevant constants) according to
\begin{align}\label{countsmom}
\nn \delta Z_{A}^{(\tx{SMOM})}&=\frac{m^{2}}{\mu^{2}}+\alpha\left\{\frac{13}{18}\left(\frac{2}{\epsilon}+\ln\frac{\overline{\mu}^{2}}{m^{2}}\right)-F\left(\frac{\mu^{2}}{m^{2}}\right)\right\},\\
\delta Z_{c}^{(\tx{SMOM})}&=\alpha\left\{\frac{1}{4}\left(\frac{2}{\epsilon}+\ln\frac{\overline{\mu}^{2}}{m^{2}}\right)-G\left(\frac{\mu^{2}}{m^{2}}\right)\right\},
\end{align}
see Eq.~\eqref{fieldstrength}. Observe that $\delta Z_{A}^{(\tx{SMOM})}$ contains an $O(\alpha^{0}_{s})$ term proportional to the gluon mass parameter $m^{2}$. This happens because in the SMOM scheme the tree-level contribution to the gluon polarization arising from the first, single-cross diagram in Fig.~\ref{figdiagrams}, $\Pi_{\tx{cross}}=m^{2}$, does not get cancelled by the equal and opposite mass term in the bare massive gluon propagator.

Due to the presence of the $O(\alpha^{0}_{s})$ term in $\delta Z_{A}^{(\tx{SMOM})}$, a naive application of Eq.~\eqref{anom} to the first of Eq.~\eqref{countsmom} would yield an anomalous dimension that is not finite in the limit $\epsilon\to 0$. In the SMOM scheme, in order to derive a finite $\gamma_{A}$, one must first subtract the divergences from Eq.~\eqref{countsmom} and then apply Eq.~\eqref{anom} to the resulting finite field-strength counterterms\footnote{Equivalently, one could derive the anomalous dimensions by a term-by-term matching of coefficients in the Callan-Symanzik equation for the inverse dressed propagators.}. By doing so, one obtains the following one-loop SMOM scheme anomalous dimensions:
\begin{align}\label{anomsmom}
\nn\gamma_{A}^{(\tx{SMOM})}&=-\frac{\mu^{2}}{\mu^{2}+m^{2}}\left\{\frac{m^{2}}{\mu^{2}}+\alpha\,\frac{\mu^{2}}{m^{2}}\ F'(\mu^{2}/m^{2})\right\},\\
\gamma_{c}^{(\tx{SMOM})}&=-\alpha\,\frac{\mu^{2}}{m^{2}}\ G'(\mu^{2}/m^{2}).
\end{align}
In Ref.~\cite{beta} the same result was found by direct integration of the RG flow. In the limit of large renormalization scales, using Eq.~\eqref{functhighlim}, it is easy to see that $\gamma_{A}^{(\tx{SMOM})}$ and $\gamma_{c}^{(\tx{SMOM})}$ reduce to the one-loop anomalous dimensions of ordinary PT.

The one-loop SMOM beta function can be computed from Eq.~\eqref{anomsmom} and Eq.~\eqref{beta}, yielding
\begin{align}\label{betasmom}
\nn\beta_{\alpha}^{(\tx{SMOM})}&=-\frac{\alpha m^{2}}{\mu^{2}+m^{2}}-\alpha^{2}\frac{\mu^{2}}{m^{2}}\bigg\{\frac{\mu^{2}}{\mu^{2}+m^{2}}\ F'(\mu^{2}/m^{2})+\\
&\qquad\qquad\qquad\qquad\qquad\quad\qquad+2G'(\mu^{2}/m^{2})\bigg\}.
\end{align}
As in the MOM scheme, $\beta_{\alpha}^{(\tx{SMOM})}$ explicitly depends on the renormalization scale $\mu$ and reduces to the ordinary perturbative beta function for $\mu\gg m$. At variance with $\beta^{(\tx{MOM})}_{\alpha}$, it contains an $O(\alpha_{s})$ term and a different scale-dependent pre-factor for the derivative $F'(s)$.\\

\begin{figure}[t]
\vskip 1cm
\centering
\includegraphics[width=0.30\textwidth,angle=-90]{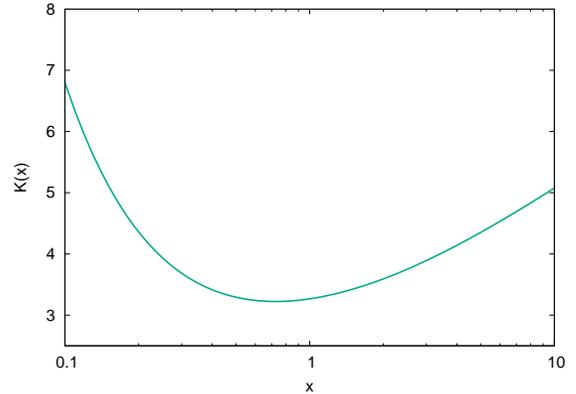}
\caption{Function $K(x)$. The minimum $K(x_{0})\approx 3.224$ is found at $x_{0}\approx0.726$.}
\label{figkfunct}
\end{figure}

The differential equation for the one-loop SMOM running coupling reads
\BE\label{diffsmom}
\frac{d\alpha^{(\tx{SMOM})}}{ds}=-b_{-1}\,\alpha^{(\tx{SMOM})}-b_{0}\,\left(\alpha^{(\tx{SMOM})}\right)^{2},
\EE
where $s=\mu^{2}/m^{2}$ and
\begin{align}\label{b1b0}
\nn b_{-1}(s)&=\frac{1}{s(s+1)},\\
b_{0}(s)&=\left\{\frac{s}{s+1}F'(s)+2G'(s)\right\}.
\end{align}
Eq.~\eqref{diffsmom} can be integrated exactly, yielding
\begin{align}\label{smomgen}
\nn&\alpha^{(\tx{SMOM})}(s)=\\
&\qquad=\frac{\alpha^{(\tx{SMOM})}(s_{0})e^{-\int_{s_{0}}^{s}ds'b_{-1}(s')}}{1+\alpha^{(\tx{SMOM})}(s_{0})\int_{s_{0}}^{s}ds'b_{0}(s')e^{-\int_{s_{0}}^{s'}ds''b_{-1}(s'')}},
\end{align}
where $s_{0}=\mu_{0}^{2}/m^{2}$ is the initial renormalization scale. With $b_{-1}(s)$ and $b_{0}(s)$ as in Eq.~\eqref{b1b0}, we find
\begin{align}\label{integrals}
\nn\exp\left(-\int_{s_{0}}^{s}ds'b_{-1}(s')\right)&=\frac{s+1}{s}\frac{s_{0}}{s_{0}+1},\\
\int_{s_{0}}^{s}ds'b_{0}(s')e^{-\int_{s_{0}}^{s'}ds''b_{-1}(s'')}&=\frac{s_{0}}{s_{0}+1}\left[K(s)-K(s_{0})\right],
\end{align}
where the function $K(x)$, shown in Fig.~\ref{figkfunct}, is defined as\footnote{$\tx{Li}_{2}(z)$ is the dilogarithm, $\tx{Li}_{2}(z)=\sum_{n=1}^{+\infty}\frac{z^{n}}{n^{2}}$.}
\begin{align}\label{functk}
\nn K(x)&=\int dx\ \left\{H'(x)+\frac{2}{x}\,G'(x)\right\}=\\
\nn&=H(x)-\frac{1}{3}\ \bigg\{\text{Li}_{2}(-x)+\frac{1}{2}\ \ln^{2}x+\\
&\quad+\frac{x^{3}+1}{3x^{3}}\ \ln(1+x)-\frac{1}{3}\ \ln x-\frac{1}{3x^{2}}+\frac{1}{6x}\bigg\}
\end{align}
and differs from the $H(x)$ of the MOM scheme by the integral of $2G'(x)/x$, which was evaluated analytically in Eq.~\eqref{functk}. 

Using Eq.~\eqref{integrals}, the one-loop SMOM running coupling, Eq.~\eqref{smomgen}, can be brought to the final form
\begin{widetext}
\BE\label{coupsmom}
\alpha^{(\tx{SMOM})}(\mu^{2})=\frac{\mu^{2}+m^{2}}{\mu^{2}}\frac{\frac{\mu_{0}^{2}}{\mu_{0}^{2}+m^{2}}\,\alpha^{(\tx{SMOM})}(\mu_{0}^{2})}{1+\frac{\mu^{2}_{0}}{\mu^{2}_{0}+m^{2}}\,\alpha^{(\tx{SMOM})}(\mu_{0}^{2})\,\left[K(s)-K(s_{0})\right]}.
\EE
\end{widetext}

\begin{figure}[t]
\vskip 1cm
\centering
\includegraphics[width=0.30\textwidth,angle=-90]{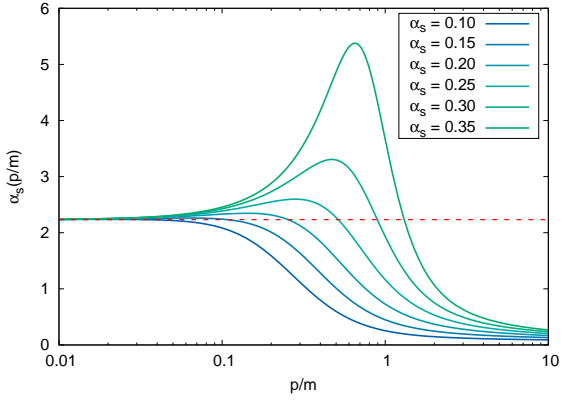}
\caption{$N=3$ one-loop running coupling of the screened expansion in the SMOM scheme for different initial values of the coupling at the scale $\mu_{0}/m=6.098$. With $m=0.656$ GeV, this corresponds to $\mu_{0}=4$ GeV. The running coupling develops a Landau pole for $\alpha_{s}^{(\tx{SMOM})}(\mu_{0}^{2})\geq 0.425$. The dashed red line displays the limiting value $\alpha_{s}^{(\tx{SMOM})}(0)\approx 2.234$.}
\label{figcoupsmom}
\end{figure}

At large renormalization scales, as long as the initial scale $\mu_{0}$ is much larger than $m$ and because of the high energy limit
\BE
K(x)\to\frac{11}{9}\ln x\qquad (x\to \infty)
\EE
(cf. the Appendix), the one-loop SMOM running coupling reduces to the standard perturbative coupling, Eq.~\eqref{coupord}. At intermediate and low momenta, on the other hand, its behavior is entirely different from that of both the ordinary PT and MOM-scheme couplings (see Fig.~\ref{figcoupsmom}).

At scales of the order of the gluon mass parameter, as in the MOM scheme, the $\mu$-dependence of the SMOM beta function is responsible for a slowing down of the running of the coupling. Indeed, due to the inequality
\BE
K(s)\geq K(\mu_{\star}^{\prime\,2}/m^{2})\approx 3.224,
\EE
where $\mu_{\star}^{\prime\,2}/m^{2}$ is the position of the minimum of $K(s)$,
\BE
\mu_{\star}'\approx0.852\ m,
\EE
$\alpha^{(\tx{SMOM})}(\mu^{2})$ does not develop a Landau pole so long as $\alpha^{(\tx{SMOM})}(\mu_{0}^{2})$ is smaller than the scale-dependent threshold value
\BE\label{threshsmom}
\alpha_{\tx{pole}}^{(\tx{SMOM})}(\mu_{0}^{2})=\frac{\mu_{0}^{2}+m^{2}}{\mu_{0}^{2}}\frac{1}{K(\mu_{0}^{2}/m^{2})-K(\mu_{\star}^{\prime\,2}/m^{2})}.
\EE
At $\mu_{0}=6.098\,m$ (corresponding to $\mu=4$ GeV in physical units), Eq.~\eqref{threshsmom} reads
\BE
\alpha_{\tx{pole}}^{(\tx{SMOM})}(6.098\,m)\approx0.304,
\EE
or, in terms of $\alpha_{s}=4\pi\alpha/3N$, for $N=3$,
\BE
\alpha_{s,\tx{pole}}^{(\tx{SMOM})}(6.098\,m)\approx0.425.
\EE
If $\alpha^{(\tx{SMOM})}(\mu_{0}^{2})<\alpha^{(\tx{SMOM})}_{\tx{pole}}(\mu_{0}^{2})$, the running coupling attains a maximum at the renormalization scale that solves the equation
\BE\label{coupsmommax}
\beta_{\alpha}^{(\tx{SMOM})}=0\ \Longleftrightarrow\ 1+\alpha^{(\tx{SMOM})}(s)s^{2}K'(s)=0.
\EE
That Eq.~\eqref{coupsmommax} always admits a solution follows from the asymptotic limits
\begin{align}
\nn 1+\alpha^{(\tx{SMOM})}(s)s^{2}K'(s)&\to-\frac{4s}{15}\ln^{2}s<0\qquad(s\to 0),\\
1+\alpha^{(\tx{SMOM})}(s)s^{2}K'(s)&\to\frac{s}{\ln s}>0\qquad(s\to \infty)
\end{align}
(cf. the Appendix). At variance with the MOM scheme and due to the pre-factor $(\mu^{2}+m^{2})/\mu^{2}$ in Eq.~\eqref{coupsmom}, the position of the maximum of the one-loop SMOM running coupling is not fixed. Instead, it depends on the initial conditions of the RG flow and shifts towards lower renormalization scales as $\alpha^{(\tx{SMOM})}(\mu_{0}^{2})$ is decreased. In the limit of very small $\alpha^{(\tx{SMOM})}(\mu_{0}^{2})$'s, an expansion of the solutions of Eq.~\eqref{coupsmommax} around $s=0$ yields
\BE
\ln^{2}s-6\,\frac{1+m^{2}/\mu_{0}^{2}}{\alpha^{(\tx{SMOM})}(\mu_{0}^{2})}=0.
\EE
Therefore, in the limit of vanishingly small initial couplings, the maximum of $\alpha^{(\tx{SMOM})}(\mu^{2})$ is attained at the scale
\BE
\mu=m\exp\left(-\sqrt{\frac{3}{2}\frac{1+m^{2}/\mu_{0}^{2}}{\alpha^{(\tx{SMOM})}(\mu_{0}^{2})}}\right).
\EE
Being its position exponentially suppressed, for small enough initial values of the coupling the maximum is essentially indistinguishable from the $\mu\to 0$ limit of $\alpha^{(\tx{SMOM})}(\mu^{2})$. The latter reads
\BE
\alpha^{(\tx{SMOM})}(\mu^{2})\to\frac{8}{5}\left\{1+\frac{4}{15}\frac{\mu^{2}}{m^{2}}\ln^{2}(\mu^{2}/m^{2})\right\}\quad(\mu\to 0),
\EE
so that the one-loop SMOM coupling saturates to a finite value, given in terms of $\alpha_{s}$ by
\BE
\alpha_{s}^{(\tx{SMOM})}(0)=\frac{32\pi}{15N}\approx2.234
\EE
for $N=3$.
\\

The one-loop SMOM RG-improved propagators are readily derived from Eqs.~\eqref{rgprop}, \eqref{rgpref} and \eqref{condsmom}. With
\begin{align}
\nn\widehat{\Delta}^{(\tx{SMOM})}(p^{2})&=\frac{1}{p^{2}+m^{2}},\\
\widehat{\mc{G}}^{(\tx{SMOM})}(p^{2})&=-\frac{1}{p^{2}},
\end{align}
we find that, when renormalized at the scale $\mu_{0}$,
\begin{widetext}
\begin{align}\label{rgpropsmom}
\nn\Delta^{(\tx{SMOM})}(p^{2};\mu_{0}^{2})&=\frac{1}{p^{2}+m^{2}}\exp\left(-\int_{\mu_{0}^{2}/m^{2}}^{p^{2}/m^{2}}ds\ \frac{1}{s+1}\left\{\frac{1}{s}+\alpha^{(\tx{SMOM})}(s)\,sF'(s)\right\}\right),\\
\mc{G}^{(\tx{SMOM})}(p^{2};\mu_{0}^{2})&=-\frac{1}{p^{2}}\exp\left(-\int_{\mu_{0}^{2}/m^{2}}^{p^{2}/m^{2}}ds\ \alpha^{(\tx{SMOM})}(s)\,G'(s)\right).
\end{align}
Equivalently, the first of Eq.~\eqref{rgpropsmom} can be expressed as
\BE
\Delta^{(\tx{SMOM})}(p^{2};\mu_{0}^{2})=\frac{1}{p^{2}}\frac{\mu_{0}^{2}}{\mu_{0}^{2}+m^{2}}\exp\left(-\int_{\mu_{0}^{2}/m^{2}}^{p^{2}/m^{2}}ds\ \frac{s}{s+1}\,\alpha^{(\tx{SMOM})}(s)\,F'(s)\right).
\EE
\end{widetext}

\begin{figure}[b]
\vskip 1cm
\centering
\includegraphics[width=0.30\textwidth,angle=-90]{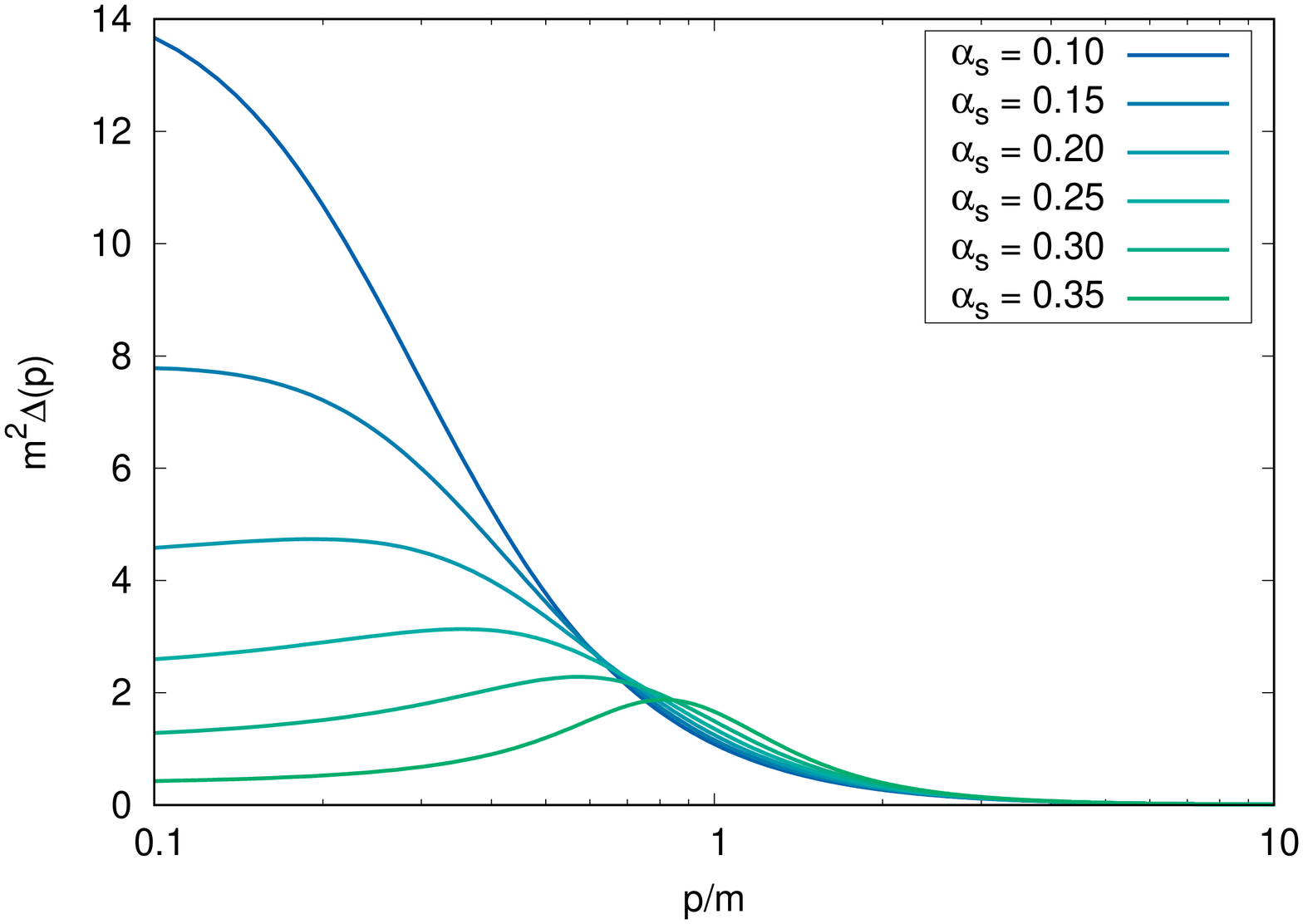}
\caption{$N=3$ one-loop RG-improved gluon propagator in the SMOM scheme, renormalized at the scale $\mu_{0}/m=6.098$ (corresponding to $\mu_{0}=4$ GeV for $m=0.656$ GeV), computed for different initial values of the coupling at the same scale.}
\label{figglusmom}
\end{figure}

\begin{figure}[t]
\vskip 1cm
\centering
\includegraphics[width=0.30\textwidth,angle=-90]{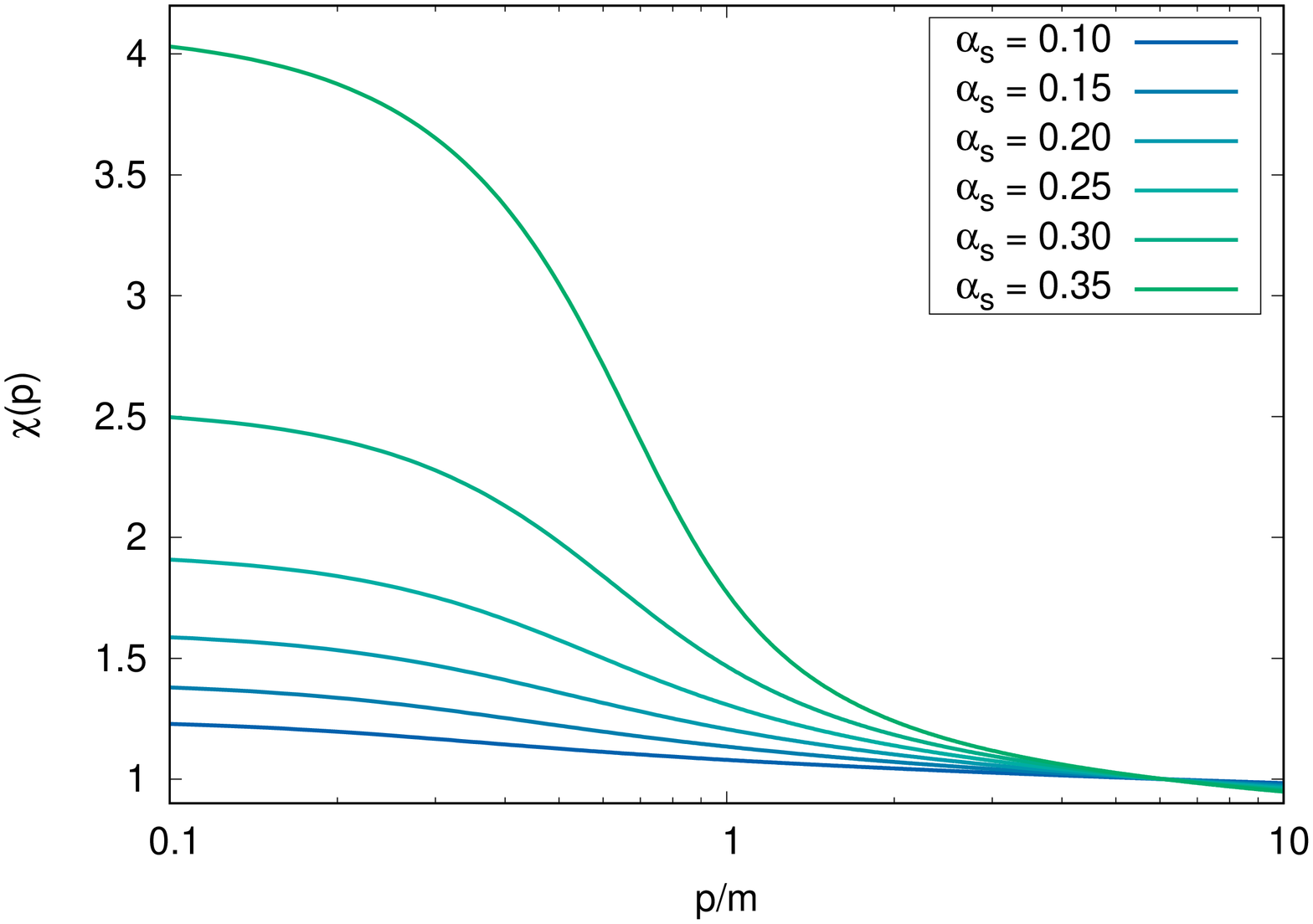}
\caption{$N=3$ one-loop RG-improved ghost dressing function $\chi(p)=-p^{2}\mc{G}(p)$ in the SMOM scheme, renormalized at the scale $\mu_{0}/m=6.098$ (corresponding to $\mu_{0}=4$ GeV for $m=0.656$ GeV), computed for different initial values of the coupling at the same scale.}
\label{figghosmom}
\end{figure}

The improved gluon propagator and ghost dressing function renormalized at the scale $\mu_{0}=6.098\, m$ (corresponding to $\mu_{0}=4$ GeV in physical units) are shown in Figs. ~\ref{figglusmom} and \ref{figghosmom}, respectively, for different initial values of the coupling constant below the threshold value $\alpha_{s,\tx{pole}}^{(\tx{SMOM})}\approx 0.43$. In the high momentum limit both the SMOM anomalous dimensions and running coupling reduce to the respective standard one-loop expressions. Therefore, Eq.~\eqref{stdrgprop} is also verified in the SMOM scheme for $p,\mu_{0}\gg m$. At intermediate and low momenta, the general behavior of the SMOM propagators parallels that of the MOM scheme. In particular, provided that the SMOM running coupling does not develop a Landau pole, the gluon propagator attains a maximum at the momentum $p=\sqrt{s}\,m$ that solves the equation
\BE\label{glumaxsmom}
1+\frac{s^{2}}{s+1}\alpha^{(\tx{SMOM})}(s)F'(s)=0.
\EE
Eq.~\eqref{glumaxsmom} always admits a solution, since
\begin{align}
\nn1+\frac{s^{2}}{s+1}\,\alpha^{(\tx{SMOM})}(s)F'(s)&\to-\frac{4}{15}\,s\ln^{2}s\leq 0\quad(s\to 0),\\
1+\frac{s^{2}}{s+1}\,\alpha^{(\tx{SMOM})}(s)F'(s)&\to1>0\quad(s\to \infty)
\end{align}
(cf. the Appendix). As in the MOM scheme, the position of the maximum depends on the initial conditions of the RG flow and shifts to lower momenta as $\alpha^{(\tx{SMOM})}(\mu_{0}^{2})$ is decreased. In the limit of vanishing momenta, since for $s\to 0$
\begin{align}
\nn \frac{s}{s+1}\,\alpha^{(\tx{SMOM})}(s)\,F'(s)&\to-\frac{1}{s},\\
\alpha^{(\tx{SMOM})}(s)\,G'(s)&\to-\frac{4}{15}\,\ln s
\end{align}
(cf. the Appendix), the one-loop improved propagators again have the same behavior as in the MOM scheme, Eq.~\eqref{propmomlow}. In particular, while the ghosts remain massless, the gluons acquire a mass.

\

In the SMOM scheme, the one-loop running coupling has a distinctive behavior: as we saw, after attaining a maximum at an intermediate scale, at low momenta it saturates to a finite value which does not depend on the initial conditions of the RG flow, namely $\alpha_{s}^{(\tx{SMOM})}(0)\approx 2.23$ (for $N=3$). As a consequence, regardless of the initial conditions, in the whole range $\mu\lesssim m$ the values of the one-loop SMOM running coupling become quite large for the perturbative standards. We should then expect the higher orders of the perturbative series to become non-negligible at scales lower than $m$. The situation is somewhat worse than in the MOM scheme: in the latter, the one-loop running coupling at any fixed scale is an increasing function of $\alpha_{s}^{(\tx{MOM})}(\mu_{0}^{2})$, so that, at least in principle, for sufficiently small initial values of the coupling the one-loop results can still provide a good approximation to the exact propagators if the gluon mass parameter $m$ is chosen appropriately. In the SMOM scheme, on the other hand, it is the fixed value of the zero-momentum coupling that dominates over the low-energy behavior of $\alpha_{s}^{(\tx{SMOM})}(\mu^{2})$. In particular, we should expect the perturbative series to converge more slowly in the SMOM scheme, rather than in the MOM scheme.

\subsection{Comparison between the MOM and the SMOM schemes}

As shown in Secs. IIIA and IIIB, both the MOM and the SMOM one-loop running coupling and RG-improved propagators have the ordinary perturbative UV limit. In the IR, the behavior of the propagators is in mutual qualitative agreement, while that of the running couplings shows significant differences. In order to make a quantitative comparison between the predictions of the two schemes, what we need to do is find a correspondence between the values of their renormalized couplings.

\begin{figure}[t]
\vskip 1cm
\centering
\includegraphics[width=0.30\textwidth,angle=-90]{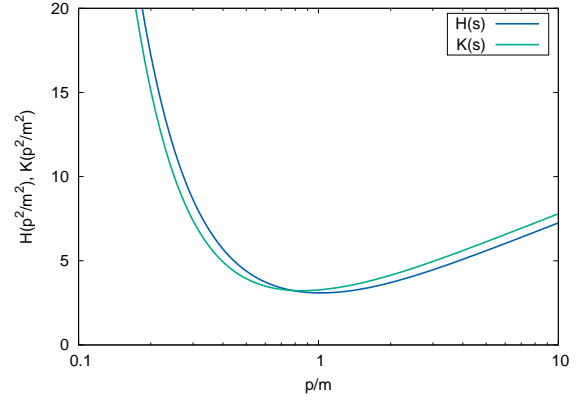}
\caption{$H(s)$ and $K(s)$ as functions of the ratio $p/m$.}
\label{figfuncts}
\end{figure}

The qualitative difference between the MOM and the SMOM one-loop running couplings ultimately originates in the pre-factor $(\mu^{2}+m^{2})/\mu^{2}$ in Eq.~\eqref{coupsmom}. Indeed, if we define a function $\widetilde{\alpha}^{(\tx{SMOM})}(\mu^{2})$ such that
\BE\label{smommod}
\alpha^{(\tx{SMOM})}(\mu^{2})=\frac{\mu^{2}+m^{2}}{\mu^{2}}\,\widetilde{\alpha}^{(\tx{SMOM})}(\mu^{2}),
\EE
then
\BE\label{coupsmomtil}
\widetilde{\alpha}^{(\tx{SMOM})}(\mu^{2})=\frac{\widetilde{\alpha}^{(\tx{SMOM})}(\mu_{0}^{2})}{1+\widetilde{\alpha}^{(\tx{SMOM})}(\mu_{0}^{2})\left[K(s)-K(s_{0})\right]}
\EE
is formally identical to the MOM running coupling, Eq.~\eqref{coupmom}, with the substitution $H(s)\to K(s)$. As shown in Fig.~\ref{figfuncts}, the functions $H(s)$ and $K(s)$ themselves have the same qualitative behavior.

The factor $(\mu^{2}+m^{2})/\mu^{2}$ in Eq.~\eqref{smommod} is a by-product of the $O(\alpha_{s}^{0})$ term in the SMOM gluon anomalous dimension, Eq.~\eqref{anomsmom}, which results in the SMOM beta function $\beta_{\alpha}^{(\tx{SMOM})}$ containing an $O(\alpha_{s})$ term. This is made explicit by computing the beta function analogue associated to $\widetilde{\alpha}^{(\tx{SMOM})}(\mu^{2})$: to one loop
\BE
\beta_{\widetilde{\alpha}}^{(\tx{SMOM})}=\frac{d\widetilde{\alpha}^{(\tx{SMOM})}}{d\ln\mu^{2}}=-\left(\widetilde{\alpha}^{(\tx{SMOM})}\right)^{2}\,\frac{\mu^{2}}{m^{2}}\ K'\left(\frac{\mu^{2}}{m^{2}}\right).
\EE
The latter contains no $O(\alpha_{s}^{0})$ terms and has the same form of the MOM beta function, Eq.~\eqref{betamom}, again with the substitution $H(s)\to K(s)$. At the level of the renormalization conditions that define the two schemes, the appearance of the factor of $(\mu^{2}+m^{2})/\mu^{2}$ can be understood as follows. From Eq.~\eqref{couptaylor} we know that in the Taylor scheme\\
\BE\label{couplingratio}
\frac{\alpha^{(\tx{SMOM})}(\mu^{2})}{\alpha^{(\tx{MOM})}(\mu^{2})}=\frac{Z_{A}^{(\tx{SMOM})}(\mu^{2})\left(Z_{c}^{(\tx{SMOM})}(\mu^{2})\right)^{2}}{Z_{A}^{(\tx{MOM})}(\mu^{2})\left(Z_{c}^{(\tx{MOM})}(\mu^{2})\right)^{2}}.
\EE
Now, while $Z_{c}^{(\tx{SMOM})}$, $Z_{A}^{(\tx{MOM})}$ and $Z_{c}^{(\tx{MOM})}$ are all equal to $1$ to $O(\alpha_{s}^{0})$,
\BE
Z_{A}^{(\tx{SMOM})}(\mu^{2})=1+\frac{m^{2}}{\mu^{2}}+O(\alpha_{s}).
\EE
Therefore
\BE\label{conversion0}
\frac{\alpha^{(\tx{SMOM})}(\mu^{2})}{\alpha^{(\tx{MOM})}(\mu^{2})}=\frac{\mu^{2}+m^{2}}{\mu^{2}}+O(\alpha_{s}).
\EE
In the next section we will show that the relation $\alpha^{(\tx{SMOM})}(\mu^{2})=(\mu^{2}+m^{2})/\mu^{2}\,\times\,\alpha^{(\tx{MOM})}(\mu^{2})$ is indeed exact, although not necessarily satisfied at any finite order in perturbation theory.

In conclusion, we find that the conversion factor between $\alpha^{(\tx{SMOM})}$ and $\alpha^{(\tx{MOM})}$ is precisely $(\mu^{2}+m^{2})/\mu^{2}$: in order to compare the two schemes, to one loop we need to choose values of the couplings such that $\alpha^{(\tx{MOM})}(\mu^{2}_{0})=\widetilde{\alpha}^{(\tx{SMOM})}(\mu^{2}_{0})$. At $\mu_{0}=6.098\,m$ (corresponding to $4$ GeV in physical units), this translates into
\BE
\alpha^{(\tx{SMOM})}(\mu_{0}^{2})\approx 1.027\, \alpha^{(\tx{MOM})}(\mu_{0}^{2}).
\EE

\begin{figure}[t]
\vskip 1cm
\centering
\includegraphics[width=0.30\textwidth,angle=-90]{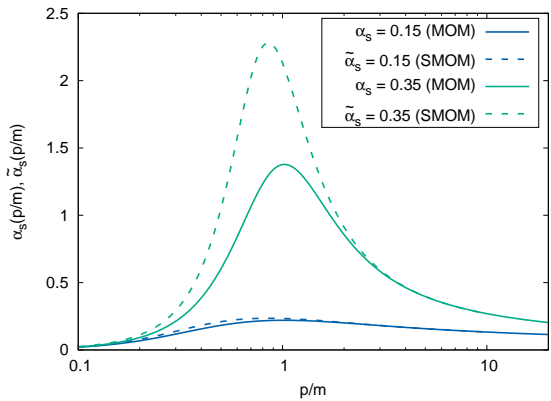}
\caption{Comparison between the $N=3$ MOM and SMOM one-loop running couplings renormalized at the scale $\mu_{0}/m=6.098$ (corresponding to $\mu_{0}=4$ GeV for $m=0.656$ GeV). For $N=3$, the MOM running coupling develops a Landau pole at $\alpha_{s}^{(\tx{MOM})}(\mu_{0}^{2})\approx0.469$, while the SMOM running coupling develops it at $\widetilde{\alpha}_{s}^{(\tx{SMOM})}(\mu_{0}^{2})\approx0.413$. See the text for the details of the comparison.}
\label{figcoupcomp}
\end{figure}
\

For our first comparison, in Fig.~\ref{figcoupcomp} we show the one-loop MOM and SMOM running couplings for two different values of $\alpha_{s}$ at the initial renormalization scale $\mu_{0}=6.098\, m$. The SMOM coupling is plotted in terms of $\widetilde{\alpha}^{(\tx{SMOM})}_{s}$, as per Eq.~\eqref{conversion0}. As discussed above, $\alpha^{(\tx{MOM})}(\mu^{2})$ and $\widetilde{\alpha}^{(\tx{SMOM})}(\mu^{2})$ have the same qualitative behavior: they both attain a maximum at a fixed scale of the order of $m$ and tend to zero at vanishing renormalization scales. The position of the maximum of $\widetilde{\alpha}^{(\tx{SMOM})}_{s}(\mu^{2})$, however, lies below that of the MOM running coupling; moreover, in the whole range $p\lesssim m$ the values of $\widetilde{\alpha}^{(\tx{SMOM})}_{s}(\mu^{2})$ are generally larger than those of $\alpha^{(\tx{MOM})}_{s}(\mu^{2})$. Since $(\mu^{2}+m^{2})/\mu^{2}>1$, we find that in the IR $\alpha^{(\tx{SMOM})}_{s}(\mu^{2})>\alpha^{(\tx{MOM})}_{s}(\mu^{2})$, enforcing the idea that the SMOM perturbative series may converge more slowly than that of the MOM scheme.

\begin{figure}[b]
\vskip 1cm
\centering
\includegraphics[width=0.30\textwidth,angle=-90]{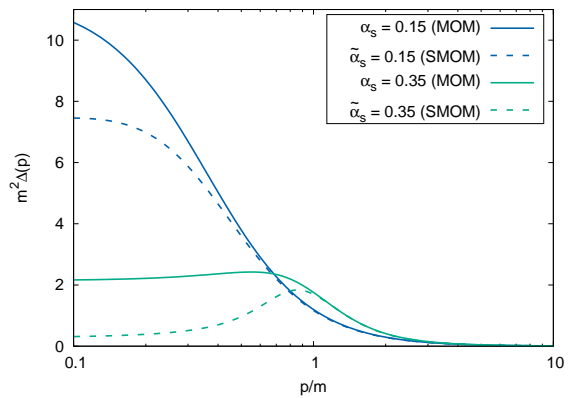}
\caption{$N=3$ one-loop RG-improved gluon propagator in the SMOM scheme, renormalized at the scale $\mu_{0}/m=6.098$ (corresponding to $\mu_{0}=4$ GeV for $m=0.656$ GeV), computed for different initial values of the coupling at the same scale.}
\label{figglucomp}
\end{figure}

\begin{figure}[t]
\vskip 1cm
\centering
\includegraphics[width=0.30\textwidth,angle=-90]{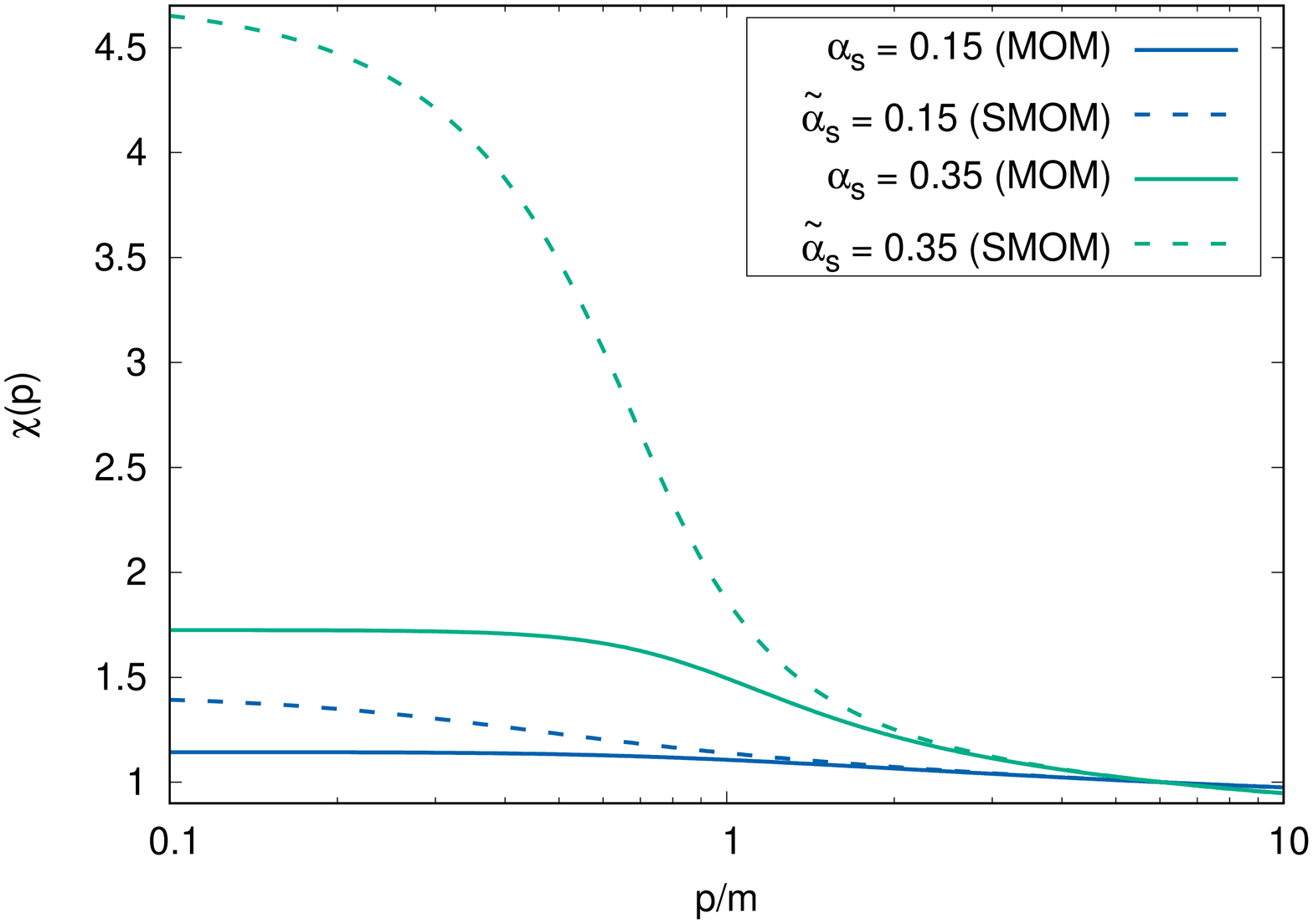}
\caption{$N=3$ one-loop RG-improved ghost dressing function $\chi(p)=-p^{2}\mc{G}(p)$ in the SMOM scheme, renormalized at the scale $\mu_{0}/m=6.098$ (corresponding to $\mu_{0}=4$ GeV for $m=0.656$ GeV), computed for different initial values of the coupling at the same scale.}
\label{figghocomp}
\end{figure}

In Figs.~\ref{figglucomp} and \ref{figghocomp} we compare the one-loop improved gluon propagators and ghost dressing functions renormalized at the scale $\mu_{0}=6.098\,m$ (corresponding to $\mu_{0}=4$ GeV in physical units) in the two schemes, with the correspondence between the renormalized couplings as discussed above. As we can see, at low momenta the propagators agree only qualitatively: at scales less than $\approx m$ the MOM gluon propagator is enhanced with respect to the SMOM propagator, while the ghost dressing function shows the opposite behavior. The relative difference between the propagators increases with the value of the coupling at $\mu_{0}$ and decreases as a function of momentum (indeed, we know that the propagators have the same, standard perturbative UV behavior in both the renormalization schemes). In the IR and for large values of the renormalized couplings the difference between the two schemes can become quite large.

\section{Optimized RG improvement and comparison with the lattice data}

By removing the Landau pole from the running of the coupling constant, the RG-improved screened massive expansion provides us with a consistent analytical framework for computing quantities at all scales in pure Yang-Mills theory, albeit at the cost of introducing a new free parameter, namely the gluon mass parameter $m$. Already at one loop, the RG-improved gluon and ghost propagators derived in such a framework display the correct qualitative behavior (as found, for example, on the lattice), being able to encode both the IR phenomenon of dynamical mass generation for the gluons and the correct UV asymptotic limits of standard perturbation theory.

Nevertheless, as discussed in Sec.~III, the one-loop RG-improved results are not expected to be quantitatively reliable below scales of the order of the gluon mass parameter $m$, the reason being that the one-loop running coupling of the screened expansion either attains a maximum at $\mu\sim m$ (in the MOM scheme) or saturates to a finite value at scales $\mu\lesssim m$ (in the SMOM scheme), becoming too large to justify the truncation of the perturbative series to first order in the coupling. In the IR, it is the one-loop fixed-scale optimized screened expansion of Refs.~\cite{xigauge,beta} that proves successful in reproducing the lattice data for the propagators: in Ref.~\cite{xigauge,beta} it was shown that the renormalization scheme in which the pole structure of the gluon propagator is gauge-invariant also yields propagators for which the terms of $O(\alpha_{s}^{2})$ and higher are negligible at low energies. The fixed-scale expansion is predictive in that its only free parameter is the energy scale of the theory, which enters the equations through the gluon mass parameter $m$ itself. We then find ourselves in possession of two distinct computational frameworks, one of which (the fixed-scale expansion) works well in the IR, while the other (the RG-improved expansion) works well in the UV. In the respective domains of applicability, both of them yield satisfactory approximations (at this stage at least qualitatively, as far as the RG-improved one is concerned) already at one loop.

A natural question to ask is whether the predictions of the two frameworks agree over some intermediate range of momenta. In general, this may depend on which values are chosen for the free parameters of the theory. Indeed, we reiterate that whereas the results of the fixed-scale expansion are completely determined once the energy scale is set by the gluon mass parameter $m$ (see Ref.~\cite{xigauge}), those of the RG-improved expansion also depend on the value of the strong coupling constant at the initial renormalization scale, $\alpha_{s}(\mu_{0}^{2})$.

Actually, the fact that in the RG-improved formalism the mass parameter $m$ and the renormalized coupling $\alpha_{s} (\mu^{2}_{0})$ can be chosen independently of one another is a major weakness of the method: already in standard perturbation theory, once the energy scale is set by the Yang-Mills analogue of $\Lambda_{\tx{QCD}}$ -- which we denote by $\Lambda_{\tx{YM}}$~--, the value of the coupling is fixed at all renormalization scales by the equation
\BE
\alpha_{s}(\mu^{2})=\frac{12\pi}{11N\ln(\mu^{2}/\Lambda_{\tx{YM}}^{2})}
\EE
(valid to one loop); in the fixed-scale framework the redundancy of free parameters is dealt with by optimization; in the formulation of the RG-improved screened PT presented in Sec.~III no such constraint exists, resulting in a loss of predictivity of the method.

The condition that the propagators and/or the running coupling computed in the fixed-scale and RG-improved frameworks match at intermediate energies can however be exploited as a criterion for fixing the value of $\alpha_{s}(\mu_{0}^{2})$: if the matching singled out a value of the coupling $\alpha_{s}(\mu_{0}^{2})$ for which the predictions of the two frameworks are in better agreement, then the gluon mass parameter $m$ -- by setting the scale for the dimensionful value of $\mu_{0}$ -- would play the same role as the $\Lambda_{\tx{YM}}$ of ordinary perturbation theory. In particular, given some value of $m$, the value of $\alpha_{s}(\mu^{2})$ at any renormalization scale would be completely determined, just as it happens in standard perturbation theory once $\Lambda_{\tx{YM}}$ is fixed. In turn, the redundancy in the free parameters of the RG-improved framework would be removed and the predictivity of the method would be restored.

In Sec.~IVA we will show that, at least in the MOM scheme, an optimal value of $\alpha_{s}(\mu_{0}^{2})$ for the matching of the fixed-scale and the RG-improved results at intermediate scales indeed exists. The predictions that follow, with the low energy behavior dictated by the fixed-scale expansion, are collected under the name of \textit{optimized} RG-improved screened PT and turn out to reproduce the lattice data quite well in the whole available range of momenta, given an appropriate choice of the energy units (cf. Sec.~IVB, where our results are compared with the data of Ref.~\cite{duarte}).

\subsection{Intermediate-scale matching of the fixed-scale and RG-improved results}

In order to determine which value of $\alpha_{s}(\mu_{0}^{2})$, if any, results in the best agreement between the IR fixed-scale and the UV RG-improved predictions, we may investigate the intermediate energy behavior either of the propagators or of the strong running coupling. In what follows we choose to work with the latter, the reason being that in the Taylor scheme the running coupling contains immediate information about both the gluon and the ghost propagators: from Eq.~\eqref{couptaylor} one finds that
\BE\label{alpharen}
\alpha_{s}(p^{2})=\alpha_{s}(\mu_{0}^{2})\,\frac{Z_{A}(p^{2})\,Z_{c}^{2}(p^{2})}{Z_{A}(\mu_{0}^{2})\,Z_{c}^{2}(\mu_{0}^{2})},
\EE
where the renormalization factors $Z_{A}(\mu^{2})$ and $Z_{c}(\mu^{2})$ can be obtained from the propagators through the relations
\BE\label{renprop}
Z_{A}(\mu^{2})=\frac{J_{B}(q^{2})}{J(q^{2};\mu^{2})}\ ,\qquad Z_{c}(\mu^{2})=\frac{\chi_{B}(q^{2})}{\chi(q^{2};\mu^{2})},
\EE
with $J(q^{2};\mu^{2})$ and $\chi(q^{2};\mu^{2})$ the gluon and ghost dressing functions renormalized at the scale $\mu^{2}$,
\begin{align}
J(q^{2};\mu^{2})&=q^{2}\,\Delta(q^{2};\mu^{2}),\nn\\
\chi(q^{2};\mu^{2})&=-q^{2}\,\mc{G}(q^{2};\mu^{2}),
\end{align}
and $J_{B}(q^{2})$ and $\chi_{B}(q^{2})$ their bare counterparts,
\begin{align}
J_{B}(q^{2})&=q^{2}\,\Delta_{B}(q^{2}),\nn\\
\chi_{B}(q^{2})&=-q^{2}\,\mc{G}_{B}(q^{2}).
\end{align}
Plugging Eqs.~\eqref{renprop} into Eq.~\eqref{alpharen} after setting $q^{2}=p^{2}$ yields the following expression for the Taylor-scheme running coupling in terms of the \textit{renormalized} gluon and ghost dressing functions:
\BE\label{alphadress}
\alpha_{s}(p^{2})=\alpha_{s}(\mu_{0}^{2})\,\frac{J(p^{2};\mu_{0}^{2})\,\chi^{2}(p^{2};\mu_{0}^{2})}{J(p^{2};p^{2})\,\chi^{2}(p^{2};p^{2})}.
\EE
In the above equation, which can be explicitly checked for the MOM and SMOM schemes of Sec.~III, the functions $J(p^{2};p^{2})$ and $\chi(p^{2};p^{2})$ define the renormalization of the propagators. For instance, in the MOM scheme
\begin{align}
J^{(\tx{MOM})}(p^{2};p^{2})=\chi^{(\tx{MOM})}(p^{2};p^{2})=1,
\end{align}
whereas in the SMOM scheme
\begin{align}
J^{(\tx{SMOM})}(p^{2};p^{2})&=\frac{p^{2}}{p^{2}+m^{2}},\nn\\
\chi^{(\tx{SMOM})}(p^{2};p^{2})&=1.
\end{align}
Apart from these functions, Eq.~\eqref{alphadress} shows that in the Taylor scheme the running coupling is proportional to a product of the gluon and ghost dressing functions, so that a comparison between the couplings of different frameworks also yields a comparison between the propagators.

Incidentally, Eq.~\eqref{renprop} can be used to prove that Eq.~\eqref{conversion0} is exact: taking the ratio between the field-strength renormalization factors defined in the SMOM and in the MOM scheme and setting $q^{2}=\mu^{2}$, we find
\BE
\frac{Z_{A}^{(\tx{SMOM})}(\mu^{2})}{Z_{A}^{(\tx{MOM})}(\mu^{2})}=\frac{\mu^{2}+m^{2}}{\mu^{2}}\ ,\quad\quad \frac{Z_{c}^{(\tx{SMOM})}(\mu^{2})}{Z_{c}^{(\tx{MOM})}(\mu^{2})}=1\ .
\EE
Once these ratios are plugged back into Eq.~\eqref{couplingratio}, the relation $\alpha^{(\tx{SMOM})}(\mu^{2})=(\mu^{2}+m^{2})/\mu^{2}\,\times\,\alpha^{(\tx{MOM})}(\mu^{2})$ is recovered, with no higher-order contributions.

The Taylor scheme is also suitable for defining a running coupling in the context of the fixed-scale perturbation theory\footnote{In the formalism of Refs.~\cite{ptqcd,ptqcd2,scaling,analyt,xigauge,damp,varT,xighost} (see also the Appendix) the gluon and ghost propagators are expressed in an essentially coupling-independent way, so that an explicit definition of what $\alpha_{s}(p^{2})$ is in the fixed-scale framework is still required. See also Ref.~\cite{beta} for a different definition of the coupling in the SMOM scheme.}. Indeed, if we renormalize the fixed-scale propagators in a MOM-like fashion, by requiring that $J(p^{2};p^{2})$ and $\chi(p^{2};p^{2})$ be momentum-independent, then we can define a fixed-scale (FS) scheme Taylor running coupling as
\BE\label{coupfs}
\alpha_{s}^{(\tx{FS})}(p^{2})=\kappa\, J^{(\tx{FS})}(p^{2})\,\chi^{(\tx{FS})}(p^{2})^{2},
\EE
where at one loop, absorbing the multiplicative renormalization constants of the dressing functions into the adimensional constant $\kappa$,
\begin{align}\label{dressfs}
J^{(\tx{FS})}(p^{2})&=\frac{1}{F(p^{2}/m^{2})+F_{0}},\nn\\
\chi^{(\tx{FS})}(p^{2})&=\frac{1}{G(p^{2}/m^{2})+G_{0}}
\end{align}
(cf. Sec.~II and the Appendix). Of course, Eqs.~\eqref{coupfs}-\eqref{dressfs} do not fix the overall normalization of $\alpha_{s}^{(\tx{FS})}(p^{2})$, which at this stage remains undefined. The constant $\kappa$ will be determined in what follows by the matching condition.

\begin{figure}[t]
\vskip 1cm
\centering
\includegraphics[width=0.30\textwidth,angle=-90]{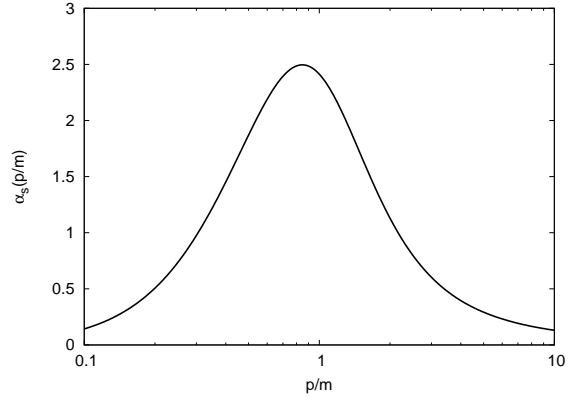}
\caption{One-loop running coupling of the screened expansion in the FS scheme. The normalization of the curve is arbitrary.}
\label{figcoupfs}
\end{figure}

The unnormalized one-loop FS running coupling is shown in Fig.~\ref{figcoupfs}. Its qualitative behavior is that of the MOM-scheme running coupling (cf. Fig.~\ref{figcoupmom}), as one would expect from having chosen momentum-independent $J(p^{2};p^{2})$ and $\chi(p^{2};p^{2})$. Accordingly, the comparison between $\alpha_{s}^{(\tx{FS})}(p^{2})$ and the SMOM running coupling will be carried out using $\widetilde{\alpha}_{s}^{(\tx{SMOM})}(p^{2})$ rather than $\alpha_{s}^{(\tx{SMOM})}(p^{2})$ (cf. the discussion in Sec.~ IIIC).

\

With $\alpha^{(\tx{FS})}_{s}(p^{2})$ as in Eq.~\eqref{coupfs} and $\alpha^{(\tx{MOM})}_{s}(p^{2})$ and $\widetilde{\alpha}^{(\tx{SMOM})}_{s}(p^{2})$ as in Eqs.~\eqref{coupmom} and \eqref{coupsmomtil}, we must now identify a range of momenta over which the running couplings of the FS and RG-improved frameworks may be expected to agree. To one loop, the latter becomes unreliable below $\mu\sim m$, corresponding to $\mu\approx 0.7$ GeV in physical units; the matching window, therefore, should lie somewhat above this value. Likewise, the upper limit of the matching interval should be set by the scale at which the one-loop results derived in the FS framework are likely to break down; this should happen at scales larger than $m$ but of the same order of $m$.

As for the normalization of the FS running coupling, under the hypothesis that at intermediate momenta the latter agrees with $\alpha^{(\tx{RG})}_{s}(p^{2})$ -- where this is taken to be either $\alpha^{(\tx{MOM})}_{s}(p^{2})$ or $\widetilde{\alpha}^{(\tx{SMOM})}_{s}(p^{2})$, depending on the scheme we are interested in --, we may require $\alpha^{(\tx{FS})}_{s}(p^{2})$ to be equal to the RG-improved coupling at some fixed renormalization scale $p=\mu_{1}$ belonging to the momentum range that we have just identified,
\BE\label{matchcond}
\alpha^{(\tx{FS})}_{s}(\mu_{1}^{2})=\alpha^{(\tx{RG})}_{s}(\mu_{1}^{2}).
\EE
This amounts to setting
\BE\label{matchscale}
\kappa=\frac{\alpha^{(\tx{RG})}_{s}(\mu_{1}^{2})}{J^{(\tx{FS})}(\mu^{2}_{1})\chi^{(\tx{FS})}(\mu^{2}_{1})^{2}}
\EE
in Eq.~\eqref{coupfs}. Of course, the actual value of the so-defined constant $\kappa$ will depend not only on the matching scale $\mu_{1}$, but also -- through $\alpha^{(\tx{RG})}_{s}(\mu_{1}^{2})$ --, on the initial value $\alpha^{(\tx{RG})}_{s}(\mu_{0}^{2})$ of the RG coupling.

\begin{figure}[t]
\vskip 1cm
\centering
\includegraphics[width=0.34\textwidth,angle=-90]{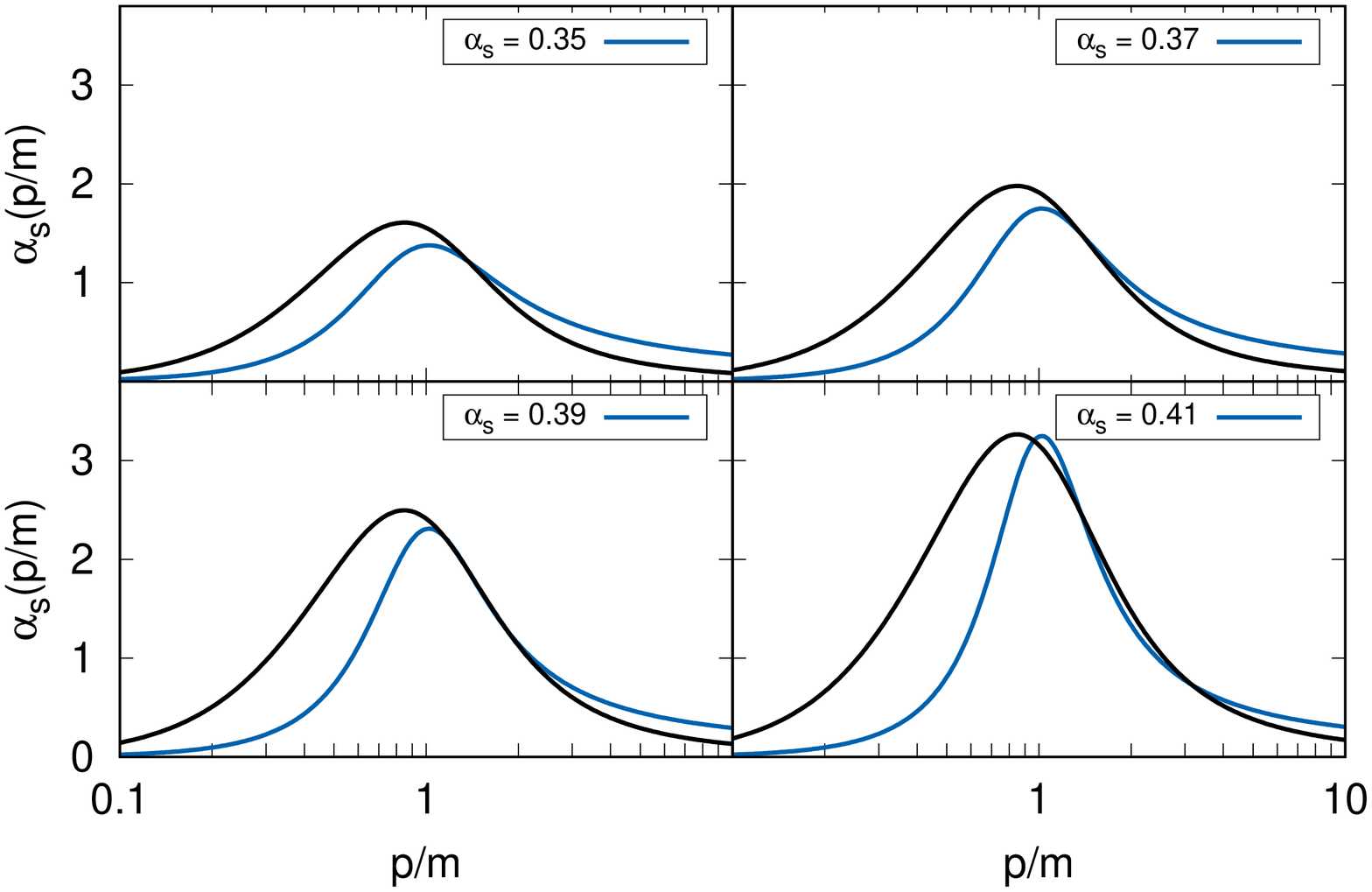}
\caption{$N=3$ intermediate-energy matching between the FS running coupling (black curves) and the MOM running coupling (blue curves) for different values of the MOM coupling renormalized at the scale $\mu_{0}/m=6.098$ (corresponding to $\mu_{0}=4$ GeV in physical units). The matching scale (see the text for details) is set to $\mu_{1}/m=1.372$ (corresponding to $\mu_{1}=0.9$~GeV).}
\label{figmatchmom}
\end{figure}
\begin{figure}[t]
\vskip 1cm
\centering
\includegraphics[width=0.34\textwidth,angle=-90]{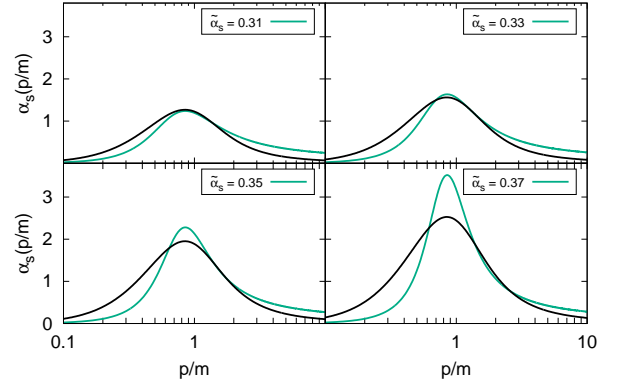}
\caption{$N=3$ intermediate-energy matching between the FS running coupling (black curves) and the SMOM running coupling (green curves) for different values of the MOM coupling renormalized at the scale $\mu_{0}/m=6.098$ (corresponding to $\mu_{0}=4$ GeV in physical units). The matching scale (see the text for details) is set to $\mu_{1}/m=1.372$ (corresponding to $\mu_{1}=0.9$~GeV).}
\label{figmatchsmom}
\end{figure}

In Figs.~\ref{figmatchmom} and \ref{figmatchsmom} we show a comparison of the normalized FS running coupling and, respectively, the MOM-scheme and SMOM-scheme running couplings, for $N=3$ and different initial values of $\alpha^{(\tx{RG})}_{s}(p^{2})$ renormalized at the scale $\mu_{0}=6.098\,m$ (corresponding to $4$ GeV in physical units). For these plots the matching scale $\mu_{1}$ was chosen equal to $1.372\,m$ (corresponding to $0.9$ GeV). Clearly, despite the $\alpha^{(\tx{RG})}_{s}(\mu_{0}^{2})$-dependent matching condition contained in Eq.~\eqref{matchscale}, the running couplings computed in the two frameworks do not agree at intermediate momenta for arbitrary values of $\alpha^{(\tx{RG})}_{s}(\mu_{0}^{2})$. In the MOM scheme, the choice $\alpha^{(\tx{MOM})}_{s}(\mu_{0}^{2})\approx 0.39$ leads to the overlap of the running couplings at scales between $p\approx m$ and $p\approx 2m$. In the SMOM scheme, on the other hand, no single choice of $\widetilde{\alpha}^{(\tx{SMOM})}_{s}(\mu_{0}^{2})$ results in the running couplings to agree over a comparably wide momentum interval\footnote{We checked that tuning the matching scale $\mu_{1}$ between $\approx m$ and $\approx 2.5\, m$ does not improve this behavior: in no case we were able to obtain an overlap between the FS and the SMOM running coupling over a wider range of momenta, without entering a regime in which the SMOM coupling develops a Landau pole.}. Why this is so can be understood in the light of the considerations made at the end of Sec.~IIIB: at scales of order $m$ and at one loop, the SMOM scheme is expected to be less reliable than the MOM scheme; therefore, under the assumption that the one-loop predictions of the FS framework are nearly exact up to $p\sim m$, the better agreement of $\alpha^{(\tx{FS})}_{s}(p^{2})$ with $\alpha^{(\tx{MOM})}_{s}(p^{2})$, rather than with $\widetilde{\alpha}^{(\tx{SMOM})}_{s}(p^{2})$, could have been anticipated. In what follows we will push no farther the comparison between the FS and the SMOM-scheme RG-improved frameworks, limiting ourselves to present our results for the MOM scheme.\\

In order to single out an optimal value of $\alpha^{(\tx{MOM})}_{s}(\mu_{0}^{2})$ for the matching, we will adopt the following criterion. Denoting with $\varepsilon(p^{2})$ the momentum-dependent relative difference between the MOM running coupling and the FS running coupling (the latter normalized as in Eq.~\eqref{matchscale}),
\BE
\varepsilon(p^{2})=\frac{\alpha^{(\tx{MOM})}_{s}(p^{2})-\alpha^{(\tx{FS})}_{s}(p^{2})}{\alpha^{(\tx{FS})}_{s}(p^{2})},
\EE
we say that $\alpha^{(\tx{MOM})}_{s}(\mu_{0}^{2})$ is optimal for the matching if it results in a MOM running coupling for which $|\varepsilon(p^{2})|\leq 1\%$ over the widest possible range of momenta in the previously identified matching interval. The matching scale $\mu_{1}$ itself -- Eq.~\eqref{matchcond} -- is fixed according to the same criterion.

\begin{figure}[t]
\vskip 1cm
\centering
\includegraphics[width=0.30\textwidth,angle=-90]{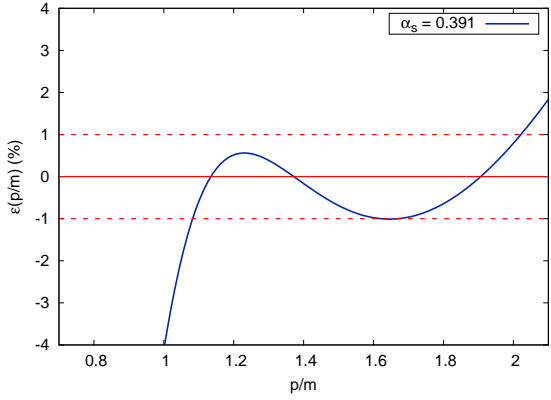}
\caption{Relative difference between the $N=3$ MOM running coupling and the FS running coupling for the optimal value $\alpha^{(\tx{MOM})}_{s}(\mu_{0}^{2})=0.391$. The initial renormalization scale is $\mu_{0}/m=6.098$ (corresponding to $\mu_{0}=4$ GeV in physical units), while the matching scale is $\mu_{1}/m=1.372$ (corresponding to $\mu_{1}=0.9$~GeV).}
\label{figmatchmomrel}
\end{figure}
\begin{figure}[t]
\vskip 1cm
\centering
\includegraphics[width=0.30\textwidth,angle=-90]{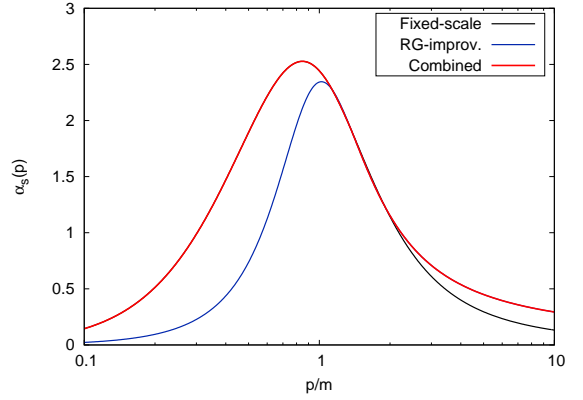}
\caption{Intermediate-energy matching between the FS running coupling (black curve) and the $N=3$ MOM running coupling (blue curve) for the optimal value $\alpha^{(\tx{MOM})}_{s}(\mu_{0}^{2})=0.391$ ($\mu_{0}=6.098\,m$, corresponding to $4$ GeV in physical units). The matching scale is $\mu_{1}=1.372\,m$ ($0.9$~GeV) and the FS coupling is normalized by $\kappa=1.200$. The red curve is obtained by combining the low-energy FS coupling and the high-energy MOM coupling.}
\label{figmatched}
\end{figure}

In Fig.~\ref{figmatchmomrel} we show the relative difference $\varepsilon(p^{2})$ computed for the optimal value $\alpha^{(\tx{MOM})}_{s}(\mu_{0}^{2})=0.391$ ($\mu_{0}=6.098\,m$, i.e. $4$ GeV in physical units), obtained for $N=3$ at the matching scale $\mu_{1}=1.372\,m$ ($0.9$ GeV) by the criterion detailed above. The range over which $|\varepsilon(p^{2})|\leq 1\%$ has width $\Delta p\approx 0.9\,m$ ($0.6$~GeV) and extends from $p\approx 1.1\,m$ to $p\approx 2\,m$. In Fig.~\ref{figmatched} the corresponding running couplings are displayed. The combined red curve, which we denote by $\alpha_{s}^{(\tx{opt})}(p^{2})$, is obtained by gluing the low-energy portion of the FS coupling to the high-energy portion of the MOM coupling at $p=\mu_{1}$. $\alpha_{s}^{(\tx{opt})}(p^{2})$ attains a maximum at $p=p_{\tx{max}}\approx 0.847\,m$ (corresponding to $0.556$ GeV in physical units),
\begin{align}
\nn p_{\tx{max}}&\approx 0.847\,m,\\
\alpha^{(\tx{opt})}_{s}(p_{\tx{max}}^{2})&\approx2.527.
\end{align}
In Sec.~IVB the combined predictions of the FS and MOM-scheme RG-improved frameworks will be compared with the lattice data for $N=3$.

\subsection{Comparison with the lattice data}

Having found that the optimal value of $\alpha^{(\tx{MOM})}_{s}(\mu_{0}^{2})$ for the matching of the $N=3$ one-loop RG-improved MOM scheme to the one-loop FS framework is $0.391$ (with $\mu_{0}=6.098\,m$ as the renormalization scale and $\mu_{1}=1.372\,m$ as the matching scale), we now proceed to compare our combined results with the lattice data of Ref.~\cite{duarte}. We reiterate that, once the RG-improved expansion is optimized by fixing $\alpha^{(\tx{MOM})}_{s}(\mu_{0}^{2})$ -- with $\mu_{0}$ expressed in units of $m$ --, the gluon mass parameter is left to stand as the only free parameter of the theory. Being a mass scale, $m$ plays the same role as $\Lambda_{\tx{YM}}$ in standard perturbation theory, entering the MOM running coupling through the ratio $p^{2}/m^{2}$ in the denominator of
\begin{align}\label{coupmommass}
\alpha_{s}^{(\tx{MOM})}(p^{2})=\frac{4\pi}{9[H(p^{2}/m^{2})-\overline{H}]}\qquad (N=3),
\end{align}
which is just Eq.~\eqref{coupmom} with $\overline{H}$ defined as
\BE
\overline{H}=H(\mu_{0}^{2}/m^{2})-\frac{4\pi}{9[\alpha_{s}^{(\tx{MOM})}(\mu_{0}^{2})]_{\tx{optim.}}}\approx2.4926
\EE
(having been obtained by optimization, $\overline{H}$ must be regarded as a constant; it does not depend neither on $m$ nor on $\mu_{0}$). As a consequence, $m$ must be inferred from experiments or, in our case, from the lattice data. Since up until this point the conversion from adimensional to physical units has been made by taking $m=0.656$ GeV (as in our previous works, see e.g. Ref.~\cite{xigauge}), in what follows we will present our results both for the aforementioned value of the mass parameter and for the value that is obtained from a fit of the combined propagators to lattice data. We remark that fitting $m$ to the lattice data only serves the purpose of fixing the energy scale of the combined results, in order to be able to compare them with the former. When all the dimensionful quantities of the theory are expressed in units of $m$, unlike the results of Sec.~III -- which still depended on a spurious free parameter --, the combined propagators are uniquely determined.

\begin{figure}[b]
\vskip 1cm
\centering
\includegraphics[width=0.30\textwidth,angle=-90]{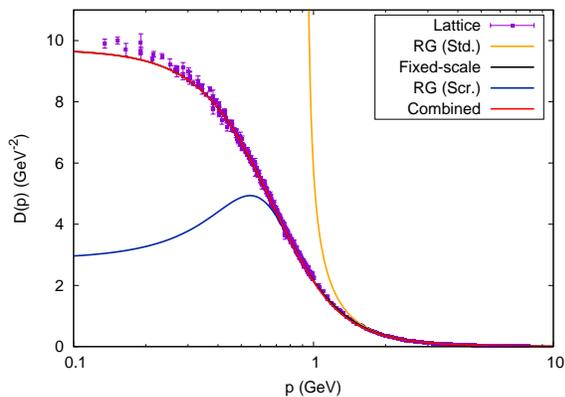}
\caption{$N=3$ gluon propagator renormalized at the scale $\mu_{0}=4$~GeV. The lattice data are taken from Ref.~\cite{duarte}. The one-loop predictions of the MOM-scheme RG-improved and FS frameworks, computed for $\alpha_{s}^{(\tx{MOM})}(\mu_{0}^{2})=0.391$ and $m=0.656$~GeV, are reported in blue and in black, respectively. The red curve is obtained by their matching at $\mu_{1}=0.9$ GeV. The orange curve is the standard perturbative one-loop RG-improved result. See the text for details.}
\label{figgluprop}
\end{figure}
\begin{figure}[t]
\vskip 1cm
\centering
\includegraphics[width=0.30\textwidth,angle=-90]{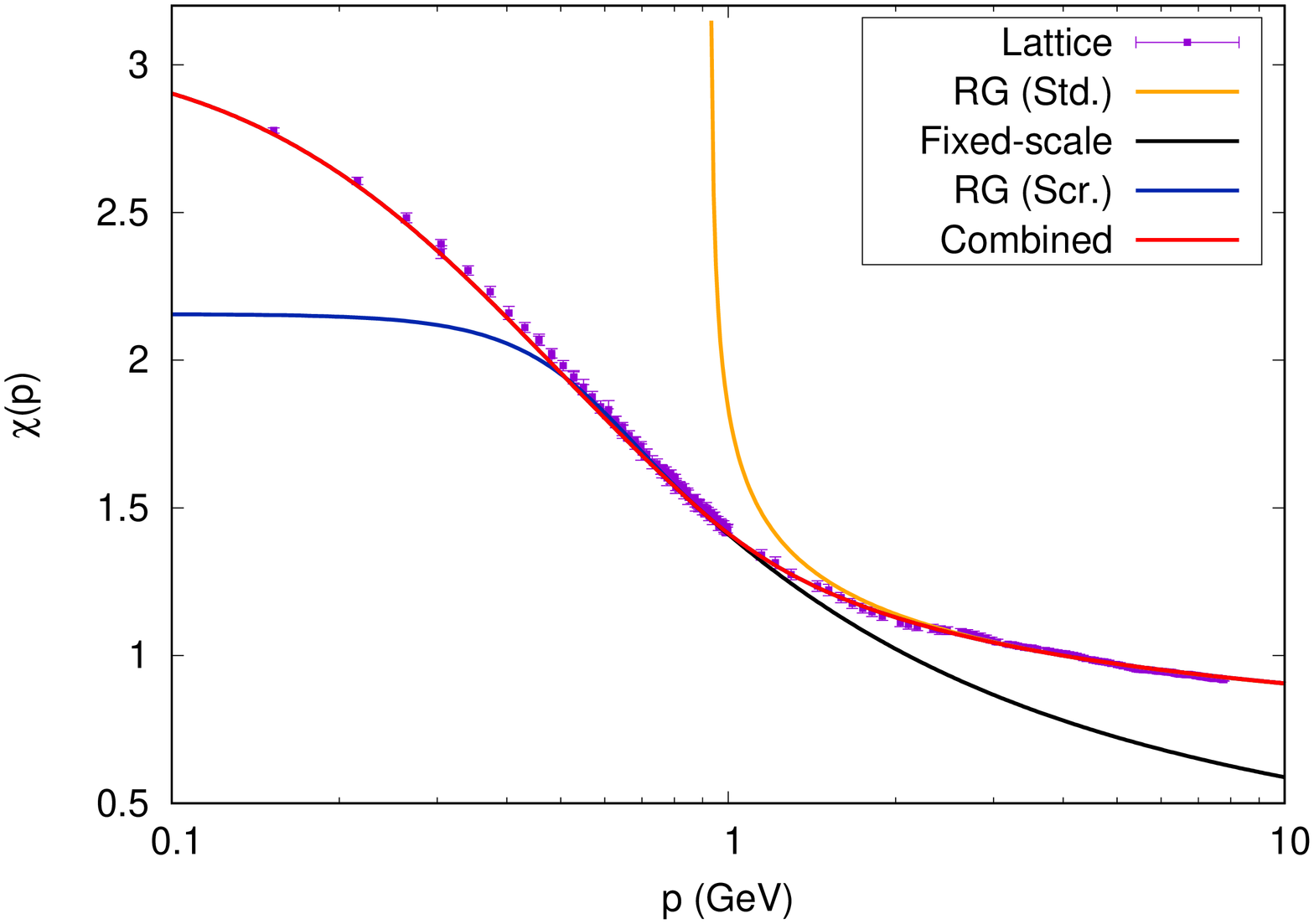}
\caption{$N=3$ ghost dressing function renormalized at the scale $\mu_{0}=4$ GeV. The lattice data are taken from Ref.~\cite{duarte}. The one-loop predictions of the MOM-scheme RG-improved and FS frameworks, computed for $\alpha_{s}^{(\tx{MOM})}(\mu_{0}^{2})=0.391$ and $m=0.656$~GeV, are reported in blue and in black, respectively. The red curve is obtained by their matching at $\mu_{1}=0.9$ GeV. The orange curve is the standard perturbative one-loop RG-improved result. See the text for details.}
\label{figghdress}
\end{figure}
\begin{figure}[t]
\vskip 1cm
\centering
\includegraphics[width=0.30\textwidth,angle=-90]{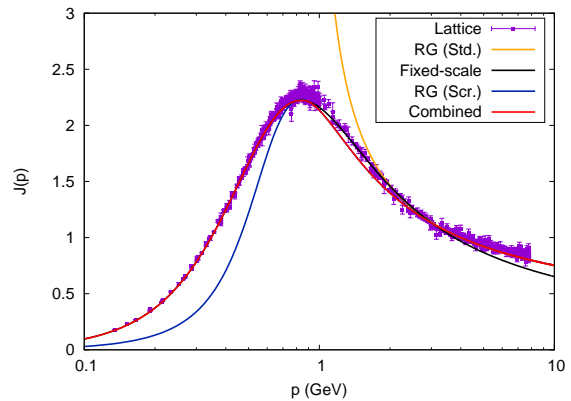}
\caption{$N=3$ gluon dressing function renormalized at the scale $\mu_{0}=4$ GeV. The lattice data are taken from Ref.~\cite{duarte}. The one-loop predictions of the MOM-scheme RG-improved and FS frameworks, computed for $\alpha_{s}^{(\tx{MOM})}(\mu_{0}^{2})=0.391$ and $m=0.656$~GeV, are reported in blue and in black, respectively. The red curve is obtained by their matching at $\mu_{1}=0.9$ GeV. The orange curve is the standard perturbative one-loop RG-improved result. See the text for details.}
\label{figgludress}
\end{figure}

In Figs.~\ref{figgluprop} and \ref{figghdress} the $N=3$ gluon propagator and ghost dressing function renormalized at the scale $\mu_{0}=4$~GeV are shown as functions of momentum. The energy scale for the analytical results is set by the gluon mass parameter $m$, preliminarly taken to be equal to $0.656$~GeV. In the figures, the red curves are obtained by combining the high-energy predictions of the RG-improved MOM scheme at $\alpha^{(\tx{MOM})}_{s}(\mu_{0}^{2})=0.391$ (displayed as blue curves) with the low-energy ones of the FS framework (displayed as black curves), the latter normalized so as to match the former at $p=\mu_{1}=0.9$ GeV. For comparison, the standard perturbative one-loop results for $\alpha_{s}(\mu_{0}^{2})=0.391$ (corresponding to $\Lambda_{\tx{YM}}=0.928$~GeV) are also displayed in the figures, as orange curves. In Fig.~\ref{figgludress} we show the $N=3$ gluon dressing functions associated to the propagators of Fig.~\ref{figgluprop}.

As we can see, already at one loop and for $m=0.656$~GeV, the combined results manage to reproduce quite well the lattice data over the whole available range of momenta (approximately 0.1~GeV to 8 GeV), especially for what concerns the ghost dressing function. At scales larger than $p\approx3$~GeV, the RG-improved screened-PT propagators are indistinguishable from their standard-PT analogues and constitute a considerable improvement over the FS screened results, which are unable to reproduce the lattice propagators for $p>1-3$~GeV. At lower, intermediate scales, as the momentum $p$ approaches $\Lambda_{\tx{YM}}$, the mass effects of screened PT kick in and the screened propagators deviate from the standard perturbative behavior, avoiding the Landau pole and following the lattice data. Below $p\approx m$, as was to be expected, the higher-order terms of the RG-improved expansion become non-negligible, and the one-loop improved MOM-scheme calculations no longer provide a good approximation to the exact results. A good approximation is nonetheless provided by the combined results, which in this regime follow the predictions of the FS framework.

\begin{figure}[b]
\vskip 1cm
\centering
\includegraphics[width=0.30\textwidth,angle=-90]{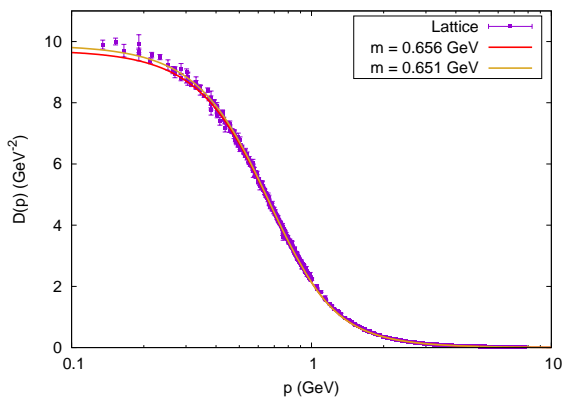}
\caption{$N=3$ gluon propagator renormalized at the scale $\mu_{0}=4$~GeV, with the lattice data of Ref.~\cite{duarte}. The one-loop predictions of the combined MOM-scheme RG-improved/FS frameworks, computed for $m=0.656$~GeV and $m=0.651$~GeV, are reported in red and gold, respectively. See the text for details.}
\label{figgluprop2}
\end{figure}
\begin{figure}[t]
\vskip 1cm
\centering
\includegraphics[width=0.30\textwidth,angle=-90]{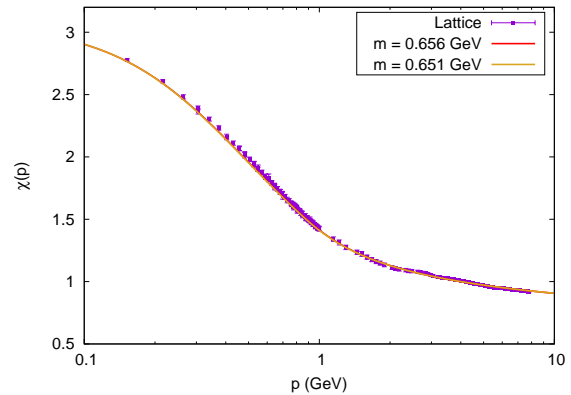}
\caption{$N=3$ ghost dressing function renormalized at the scale $\mu_{0}=4$~GeV, with the lattice data of Ref.~\cite{duarte}. The one-loop predictions of the combined MOM-scheme RG-improved/FS frameworks, computed for $m=0.656$~GeV and $m=0.651$~GeV, are reported in red and gold, respectively. See the text for details.}
\label{figghdress2}
\end{figure}
The agreement improves further if the value of $m$ is determined by fitting the combined gluon propagator to the lattice data. In Figs.~\ref{figgluprop2} and \ref{figghdress2} we show the combined gluon propagator and ghost dressing function, respectively, computed for the fitted value of the gluon mass parameter, namely $m=0.651$~GeV (the curves computed for $m=0.656$~GeV are also displayed in the figures for comparison). Clearly, the ever so slight decrease in the value of the mass parameter is sufficient to enhance the gluon propagator at low momenta, bringing it onto the lattice data without spoiling either its intermediate- and high-energy behavior, or that of the ghost dressing function.

We should remark that, for these last plots, in changing the value of $m$ the previously reported values of $\mu_{1}$ and $\mu_{0}$ in physical units have also changed. The matching scale $\mu_{1}=1.372\,m$ for combining the fixed-scale results with the MOM-scheme RG-improved ones is now equal to $0.89$~GeV (instead of 0.9~GeV, for $m=0.656$~GeV), whereas the scale $\mu_{0}=6.098\,m$, interpreted as the scale at which, by optimization, $\alpha_{s}^{(\tx{MOM})}=0.391$, now equals $3.97$~GeV (instead of 4~GeV). As for the renormalization scale of the propagators -- previously denoted also with $\mu_{0}$ and rigorously defined by Eqs.~\eqref{condmom}~--, in order to compare our results with the lattice data we had to set it back to 4~GeV, rather than keeping it equal to the new value 3.97~GeV. Indeed, observe that the scale at which the propagators are defined and the one at which the initial value of the running coupling is defined do not need to coincide, as long as the initial value of the coupling is chosen so as to follow the RG-flow. If we want to know the value of the coupling constant at 4~GeV for $m=0.651$~GeV, we can compute it directly from Eq.~\eqref{coupmommass} using physical units: we find
\BE
\alpha_{s}^{(\tx{MOM})}(4~\tx{GeV})=0.389\qquad(m=0.651\ \tx{GeV}).
\EE
Of course, the difference between $0.391$ and $0.389$, $3.97$~GeV and $4$~GeV, 0.89~GeV and 0.90~GeV, etc., is minimal; we may expect larger approximation errors to influence the numerical outcome of our analysis. Nonetheless, these calculations make explicit the role of the gluon mass parameter as the (only) mass scale of the theory, following the optimization of the screened massive expansion.

\section{Discussion}

The dynamical generation of an infrared mass for the gluons raises questions as to whether the standard expansion point of QCD perturbation theory -- namely, a massless vacuum for the gauge sector -- is an appropriate choice for describing the low-energy behavior of the theory. That in the IR a massive expansion point for the gluons may improve the QCD perturbative series is corroborated by a Gaussian Effective Potential (GEP) variational analysis of pure Yang-Mills theory: by minimizing the vacuum energy of the latter, a massive zero-order gluon propagator was shown \cite{varT} to bring us closer to the exact, non-perturbative vacuum of the gauge sector. The resulting perturbation theory -- defined by a simple shift of the kinetic and interaction Lagrangian -- was termed screened massive expansion and studied in Refs.~\cite{ptqcd,ptqcd2,scaling,analyt,xigauge,damp,varT,xighost,beta}.

In its fixed-coupling, fixed-scale formulation, the screened massive expansion proved successful in accurately reproducing the infrared lattice data for the propagators of pure Yang-Mills theory already at one loop \cite{ptqcd,ptqcd2,xigauge}. Moreover, it was proven capable of describing the phenomenon of dynamical mass generation for the gluons in a non-trivial manner: whereas the zero-order gluon propagator is massive by the definition of the method itself, the tree-level mass terms which appear in the dressed propagator cancel out, so that the saturation of the gluon propagator at zero momentum turns out to be an actual effect of the loops, i.e. of the strong interactions between the gluons. Nonetheless -- strictly speaking -- the screened expansion alone cannot be used to prove that the gluons acquire a mass in the infrared. Albeit it being a non-trivial prediction of the method for any non-zero value of the gluon mass parameter $m$, when the latter is set to zero the ordinary perturbative series of YM theory is recovered, so that no mass generation occurs. In the context of the screened expansion, that $m\neq0$ should lead to more reliable results in the IR can only be inferred from the aforementioned GEP analysis.

Following the optimization of the screened expansion by principles of gauge invariance \cite{xigauge,beta}, the gluon mass parameter $m$ is left as the only free parameter of the theory, playing the same role as the QCD/YM scale $\Lambda_{\text{YM}}$ of the standard perturbative expansion, with respect to which all the dimensionful values -- including the proper gluon's mass -- are to be measured. One could still wonder how a mass parameter, which is added and subtracted again in the Lagrangian, can have a physical role at all in the dynamics of the theory. From a variational point of view, since the optimal value of $m$ yields the best expansion around a Gaussian massive vacuum \cite{varT}, the mass parameter itself must be regarded as the best Gaussian approximation for the dynamically generated mass of the full theory. Such a mass is then subject to quantum corrections, which ultimately determine the value of the proper gluon's mass.

At energies larger than about $2$~GeV, the fixed-scale one-loop approximation breaks down due to the presence of large logarithms. This can be dealt with by resorting to ordinary RG methods, i.e. by defining a scheme-dependent running coupling constant and integrating the RG flow for the propagators. A second, most important reason to study the RG flow of the screened expansion is to address the issues related to the strong interactions' IR Landau pole. From both a theoretical and a practical point of view, the negativity of the coefficients of the standard QCD beta function (at least to five loops \cite{chetyrkin} and for a sufficiently small number of quarks), paired with the absence of mass scales in the Lagrangian (other than the quark masses), results in a strong running coupling which, in mass-independent renormalization schemes, diverges in the infrared, thus making ordinary perturbation theory inconsistent at energies of the order of the QCD scale. In order for the screened expansion to be meaningful in the IR, the Landau pole must be shown to disappear from the running coupling constant, when the former is used to compute the latter.\\

In the previous sections, the RG improvement of the screened massive expansion was studied at one loop in two renormalization schemes, namely, the MOM and the SMOM schemes, with the running coupling $\alpha_{s}(p^{2})$ defined in the Taylor scheme ($Z_{1}^{c}=1$). In both schemes, the existence of a non-perturbative mass scale set by the gluon mass parameter $m$ causes the beta function to explicitly depend on the renormalization scale, thus providing a mechanism by which the running of the coupling is allowed to slow down in the infrared. The most notable feature of the RG-improved screened expansion in the MOM and SMOM schemes is indeed the absence of Landau poles in their running couplings (at one loop and for sufficiently small initial values of the coupling), a necessary condition for the consistency of any perturbative approach which aims to be valid at all scales. Instead of diverging, the one-loop MOM running coupling $\alpha_{s}^{(\tx{MOM})}(p^{2})$ attains a maximum at the fixed scale $\mu_{\star}\approx 1.022\,m$ and then decreases to zero as $p^{2}\to 0$. The one-loop SMOM running coupling $\alpha_{s}^{(\tx{SMOM})}(p^{2})$, on the other hand, attains a maximum at a scale that depends on the initial value of the coupling, and then saturates to the finite, non-zero value $\alpha_{s}^{(\tx{SMOM})}(0)=32\pi/15N\approx2.234$ for $N=3$. Both $\alpha_{s}^{(\tx{MOM})}(p^{2})$ and $\alpha_{s}^{(\tx{SMOM})}(p^{2})$ have the ordinary perturbative (one-loop) limit in the UV, where the mass effects due to the gluon mass become negligible.

Since in both the renormalization schemes the one-loop running coupling becomes quite large at scales of the order of $m$, the one-loop predictions of the RG-improved framework are expected to become quantitatively unreliable at low energies. In particular, for comparable initial values of the coupling, the one-loop SMOM running coupling is always larger than the one-loop MOM running coupling in the IR (a feature which is mostly but not exclusively due to the saturation of the former at low momenta), so that the perturbative series is expected to converge more slowly in the SMOM scheme than in the MOM scheme.

The MOM and SMOM RG-improved gluon and ghost propagators were computed at one loop, for different initial values of the coupling constant, by numerically integrating the respective anomalous dimensions. We found that the improved propagators have the expected qualitative behavior -- as determined, for instance, by the lattice calculations --, showing mass generation for the gluons, no mass generation for the ghosts and the logarithm-to-rational-power UV tails of ordinary perturbation theory.\\

Under the hypothesis that the one-loop RG-improved results are sufficiently accurate down to $p\approx m$, the initial value of the coupling $\alpha_{s}(\mu_{0}^{2})$ -- one of the two free parameters of the RG-improved screened framework, together with the gluon mass parameter -- can be fixed by requiring the improved predictions to match those of the fixed-scale expansion at intermediate energies. The matching was found to work better in the MOM scheme, where the optimal choice $\alpha_{s}^{(\tx{MOM})}(\mu_{0}^{2})=0.391$ at $\mu_{0}=6.098\,m$ yields a running coupling which agrees to less than 1\% with its FS analogue over a momentum range of width $\Delta p\approx m$.

The optimization of the value of $\alpha_{s}(\mu_{0}^{2})$, where the initial renormalization scale $\mu_{0}$ itself is expressed in units of $m$, leaves the gluon mass parameter as the only free parameter of the RG-improved framework. This is of course highly desirable, since (modulo the renormalization conditions) pure Yang-Mills theory has only one free parameter, namely, the coupling or the QCD/YM scale $\Lambda_{\tx{YM}}$. In the optimized framework, $m$ uniquely determines the value of the running coupling at any given renormalization scale and, more generally, it sets the scale for the dimensionful values of the theory. In this sense, optimization enables us to truly regard the gluon mass parameter as the screened-expansion analogue of $\Lambda_{\tx{YM}}$.

The predictions obtained by combining the low-energy results ($p<1.372\,m$) for the propagators in the FS screened expansion with the high-energy ones ($p>1.372\,m$) of the optimized MOM-scheme RG-improved screened expansion were compared with the lattice data of Ref.~\cite{duarte} and found to be in excellent agreement if the value $m=0.651$~GeV (obtained by a fit of the data themselves) is used.\\

The intermediate-scale matching between the FS and RG-improved MOM frameworks proves to be a powerful method for quantitatively predicting the behavior of the gluon and ghost propagators, over a wide range of momenta and from first principles, already at one loop. This reinforces the idea that the full dynamics of YM theory and, perhaps, of full QCD, may be accessible by plain -- albeit optimized -- PT, by a mere change of the expansion point of the perturbative series, allowing for massive transverse gluons at tree-level.

At present, whether the optimized implementations of the screened massive expansion yield a good approximation of the exact results beyond the two-point sector remains an open issue. In this respect, it would be interesting to make use of the present formalism to study the behavior of the ghost-gluon and three-gluon vertices, which have already been computed -- for specific kinematic configurations of the external momenta -- e.g. on the lattice \cite{cucch06,cucch08c,athe,duarte2} and by the numerical integration of Schwinger-Dyson equations \cite{huber14,eich,will}. Encouraging signs that the screened expansion may work in the three-point sector come from the asymptotic analysis of the fixed-scale-framework gluon propagator $\Delta(p^{2})$ in the deep IR, where (cf. Eqs.~\eqref{propsapp} and \eqref{limsapp} in the Appendix)
\BE
Z_{\Delta}\,\Delta^{-1}(p^{2})\to\frac{5m^{2}}{8}+\frac{13}{18}\ p^{2}\ln(p^{2}/m^{2})+O(p^{2}).
\EE
Here $Z_{\Delta}$ is a multiplicative renormalization constant and the logarithmic term comes from the massless ghost loop in the gluon polarization tensor. By the Slavnov-Taylor identities, such a logarithm is inherited by the form-factor of the three-gluon vertex \cite{aguilar14b,aguilar19,aguilar20} and is responsible for its characteristic ``zero crossing'', i.e. its becoming negative at low energies, a feature which has been confirmed by multiple studies. Thus the behavior of the propagators computed in the screened expansion appears to be consistent with what we know -- both analytically and numerically -- about the three-point functions. An explicit computation of the latter will help to clarify the extent to which the screened massive expansion is able to describe the full dynamics of pure Yang-Mills theory and QCD.

\acknowledgments

This research was supported in part by ``Piano per la Ricerca di Ateneo 2017/2020 - Linea di intervento 2'' of the University of Catania.

\appendix*
\section{Fixed-scale screened PT and the functions $\boldsymbol{H(x)}$ and $\boldsymbol{K(x)}$}

In Euclidean space, the renormalized one-loop gluon polarization $\Pi_{\tx{loop}}^{(R)}$ and ghost self-energy $\Sigma_{\tx{loop}}^{(R)}$ computed in the framework of the massive screened expansion are given by \cite{ptqcd,ptqcd2}
\begin{align}
\nn\Pi_{\tx{loop}}^{(R)}(p^{2})&=-\alpha p^{2}\,(F(s)+\mathcal{C}),\\
\Sigma_{\tx{loop}}^{(R)}(p^{2})&=\alpha p^{2}\,(G(s)+\mathcal{C}'),
\end{align}
where $s=p^{2}/m^{2}$ ($m$ being the gluon mass parameter),
\BE
\alpha=\frac{3N\alpha_{s}}{4\pi}=\frac{3Ng^{2}}{16\pi^{2}},
\EE
and $\mathcal{C}$ and $\mathcal{C}'$ are renormalization-scheme-dependent constants. The adimensional functions $F$ and $G$ \cite{ptqcd,ptqcd2} are defined as
\begin{align}
\nn F(x)&=\frac{5}{8x}+\frac{1}{72}\left[L_a(x)+L_b(x)+L_c(x)+R(x)\right],\\
G(x)&=\frac{1}{12}\left[L_{g}(x)+R_{gh}(x)\right],
\label{FGx}
\end{align}
where the logarithmic functions $L_i$ are
\begin{align}
L_a(x)&=\frac{3x^3-34x^2-28x-24}{x}\>\times\nn\\
&\times\sqrt{\frac{4+x}{x}}
\ln\left(\frac{\sqrt{4+x}-\sqrt{x}}{\sqrt{4+x}+\sqrt{x}}\right),\nn\\
L_b(x)&=\frac{2(1+x)^2}{x^3}(3x^3-20x^2+11x-2)\ln(1+x),\nn\\
L_c(x)&=(2-3x^2)\ln x,\nn\\
L_{g}(x)&=\frac{(1+x)^{2}(2x-1)}{x^{2}}\ln(1+x)-2x\ln x,
\label{logsA}
\end{align}
and the rational parts $R_i$ are
\begin{align}
R(x)&=\frac{4}{x^{2}}-\frac{64}{x}+34,\nn\\
R_{gh}(x)&=\frac{1}{x}+2.
\label{rational}
\end{align}
The fixed-scale one-loop gluon and ghost propagators computed in the screened expansion can be expressed as
\begin{align}\label{propsapp}
\nn\Delta(p^{2})&=\frac{Z_{\Delta}}{p^{2}[F(p^{2}/m^{2})+F_{0}]},\\
\mc{G}(p^{2})&=-\frac{Z_{\mc{G}}}{p^{2}[G(p^{2}/m^{2})+G_{0}]},
\end{align}
where $Z_{\Delta}$ and $Z_{\mc{G}}$ are multiplicative renormalization factors and $F_{0}$ and $G_{0}$ are additive renormalization constants. In Refs.~\cite{xigauge,beta}, the latter were optimized by requirements of gauge invariance and minimal sensitivity, and their optimal value was found to be
\begin{align}
F_{0}=-0.876\,,\qquad G_{0}=0.145\,.
\end{align}
As for the functions $F$ and $G$, in the limit $x\to \infty$ we find
\begin{align}
\nn F(x)&\to\frac{13}{18}\,\ln x+\frac{17}{18}+\frac{5}{8x}+O(x^{-2}),\\
G(x)&\to\frac{1}{4}\,\ln x+\frac{1}{3}+\frac{1}{4x}+O(x^{-2}).
\end{align}
On the other hand, for $x\to 0$ \footnote{Here we correct an error in Ref.~\cite{beta}, where the coefficients of $x$ in the expansion of $L_{a}(x),\,L_{b}(x)$ and $F(x)$ around $x=0$ (Eqs.(A7)-(A8) of Ref.~\cite{beta}) were reported incorrectly.},
\begin{align}\label{limsapp}
\nn F(x)&\to\frac{5}{8x}+\frac{1}{36}\,\ln x+\frac{257}{216}+\frac{389}{1080}\,x+O(x^{2}),\\
G(x)&\to\frac{5}{24}-\frac{1}{6}\,x\ln x+\frac{2}{9}\,x+O(x^{2}).
\end{align}
\

The function $H(x)$, whose derivative is proportional to the beta function of the MOM running coupling, is defined as
\BE
H(x)=2G(x)+F(x).
\EE
For $x\to \infty$ we have
\BE
H(x)\to\frac{11}{9}\,\ln x+\frac{29}{18}+\frac{9}{8x}+O(x^{-2}),
\EE
whereas for $x\to 0$
\BE
H(x)\to\frac{5}{8x}+\frac{1}{36}\,\ln x+\frac{347}{216}-\frac{1}{3}\,x \ln x+\frac{869}{1080}\,x+O(x^{2}).
\EE
The one-loop MOM running coupling $\alpha_{s}^{(\tx{MOM})}(p^{2})$ has the following asymptotic behavior:
\BE
\alpha_{s}^{(\tx{MOM})}(p^{2})\to\frac{32\pi}{15N}\frac{p^{2}}{m^{2}}\left(1-\frac{2}{45}\frac{p^{2}}{m^{2}}\ln\frac{p^{2}}{m^{2}}\right)
\EE
as $p\to 0$ and
\BE
\alpha_{s}^{(\tx{MOM})}(p^{2})\to\frac{12\pi}{11N\ln (p^{2}/m^{2})}
\EE
as $p\to \infty$.\\

The expressions for the SMOM scheme beta function and running coupling involve the function $K(x)$, defined as
\begin{align}\label{functkx}
\nn K(x)&=\int dx\ \left\{H'(x)+\frac{2}{x}\,G'(x)\right\}=\\
\nn&=H(x)-\frac{1}{3}\ \bigg\{\text{Li}_{2}(-x)+\frac{1}{2}\ \ln^{2}x+\\
&\quad+\frac{x^{3}+1}{3x^{3}}\ \ln(1+x)-\frac{1}{3}\ \ln x-\frac{1}{3x^{2}}+\frac{1}{6x}\bigg\}
\end{align}
where $\tx{Li}_{2}(z)$ is the dilogarithm, $\tx{Li}_{2}(z)=\sum_{n=1}^{+\infty}\frac{z^{n}}{n^{2}}$. In the limit $x\to \infty$ we find
\BE
K(x)\to\frac{11}{9}\,\ln x+\frac{\pi^{2}+29}{18}+\frac{5}{8x}+O(x^{-2}),
\EE
whereas in the limit $x\to 0$
\begin{align}
\nn K(x)&\to\frac{5}{8x}-\frac{1}{6}\,\ln^{2}x+\frac{5}{36}\,\ln x+\frac{113}{72}+\\
&\quad-\frac{1}{3}\,x \ln x+\frac{1139}{1080}\,x+O(x^{2}).
\end{align}
The asymptotic limits of the one-loop SMOM running coupling $\alpha_{s}^{(\tx{SMOM})}(p^{2})$ are computed to be
\BE
\alpha_{s}^{(\tx{SMOM})}(p^{2})\to\frac{32\pi}{15N}\left(1+\frac{4}{15}\frac{p^{2}}{m^{2}}\ln^{2}\frac{p^{2}}{m^{2}}\right)
\EE
as $p\to 0$ and
\BE
\alpha_{s}^{(\tx{SMOM})}(p^{2})\to\frac{12\pi}{11N\ln (p^{2}/m^{2})}
\EE
as $p\to \infty$.

\end{document}